\documentclass[conference]{IEEEtran}

\IEEEoverridecommandlockouts
\usepackage{cite}
\usepackage{algorithmic}
\usepackage{kotex}
\usepackage{url}
\usepackage{verbatim}
\usepackage{caption}
\usepackage{gensymb}
\usepackage{amssymb}
\usepackage{amsthm}
\usepackage{multicol}
\usepackage{exscale}
\usepackage[linesnumbered,ruled,vlined]{algorithm2e}
\usepackage{listings}
\usepackage{stfloats}
\usepackage{enumitem}
\usepackage{amsmath,amssymb,amsfonts}
\usepackage{xcolor}
\usepackage{comment}
\usepackage{graphicx}
\usepackage{textcomp}
\usepackage{color}
\usepackage{setspace}
\usepackage{enumerate}
\usepackage{multirow}
\usepackage{amsmath}
\usepackage{subfig}
\usepackage{colortbl}
\usepackage{tabularx}
\usepackage{makecell}
\usepackage{array}
\usepackage[normalem]{ulem}
\usepackage{titlecaps}
\def\BibTeX{{\rm B\kern-.05em{\sc i\kern-.025em b}\kern-.08em
    T\kern-.1667em\lower.7ex\hbox{E}\kern-.125emX}}

\begin{document}

\newcommand{\cb}{\textcolor{blue}}
\newcommand{\cred}{\textcolor{red}}
\definecolor{darkviolet}{rgb}{0.58, 0.0, 0.83}
\newcommand{\cg}{\textcolor{darkviolet}}
\newcommand{\squishlist}{
	\begin{list}{$\bullet$}
		{ \setlength{\itemsep}{0pt}      \setlength{\parsep}{-0pt}
			\setlength{\topsep}{4pt}       \setlength{\partopsep}{0pt}
			\setlength{\listparindent}{-2pt}
			\setlength{\itemindent}{-5pt}
			\setlength{\leftmargin}{1em} \setlength{\labelwidth}{0em}
			\setlength{\labelsep}{0.5em} } }
	
	\newcommand{\squishend}{
\end{list}  }

\newcommand{\emap}{\textit{{{eMap}}\normalfont}}
\newcommand{\emplan}{\textit{{{eMPlan}}\normalfont}}
\newcommand{\emdyn}{\textit{{{eMDyn}}\normalfont}}
\SetKwInput{KwInput}{Input}
\SetKwInput{KwOutput}{Output}

\title{\huge Optimizing Placement of Heap Memory Objects in Energy-Constrained Hybrid Memory Systems}

\author{
	{\rm Taeuk Kim$^{\dagger}$\thanks{$^{\dagger}$Mr. Taeuk is currently affliated with Tmax Cloud, Seoul, Republic of Korea, but most of the work is done when he was in Sogang University.}, Safdar Jamil, Joongeon Park, Youngjae Kim}\\
	{\rm Dept. of Computer Science and Engineering, Soang University, Seoul, Republic of Korea}\\
	{\rm \small taeuk\_kim@tmax.co.kr \{safdar, joongeon, youkim\}@sogang.ac.kr}
}

\maketitle

\begin{abstract}
    Main memory significantly impacts the power and energy utilization of the overall server system. Non-Volatile Memory (NVM) devices, are suitable candidates for the main memory to reduce static energy consumption. But unlike DRAM, the access latencies and the dynamic energy consumption of write operation of the NVM devices are higher. Thus, Hybrid Main Memory Systems (HMMS) employing DRAM and NVM have been proposed to reduce the overall energy depletion of main memory while optimizing the performance of application. However, memory object placement is crucial for optimal performance and energy efficiency in HMMS due to high write latency and energy consumption of NVM devices. This paper proposes \emap{}, an optimal heap memory object placement planner for HMMS. \emap{} takes into account the object-level access patterns and energy consumption to provide an ideal placement policy for objects to mitigate performance and energy consumption.In particular, \emap{} is equipped with two modules,~\emplan{} and \emdyn{}. \emplan{} is a static placement planner which provides one-time placement policies for memory objects to meet the energy budget. \emdyn{} is a runtime model to consider the requests of changes in the energy constraint during the application execution. Both modules are based in Integer Linear Programming(ILP) and consider three major constraints, namely decision, capacity and energy constraints to optimally placing the memory objects in HMMS. We evaluate the proposed solution with two scientific application benchmarks, NAS Parallel Benchmark (NPB) and Problem-based Benchmark Suit (PBBS), on two testbeds by emulating the NVM using QUARTZ~\cite{quartz}. Our extensive experiments in comparison with Memory Object Classification and Allocation (MOCA) framework showed that our solution is 4.17x less costly in terms of the memory object profiling and reduce the energy consumption up to 14\% with the same performance. On the other hand, \emdyn{} module also meets the performance and energy requirement during the application execution by considering the migration cost in terms of time and energy. 
\end{abstract}


\setstretch{0.95}

\section{Introduction}
\label{sec:intro}

In the computing system, there are two major components to account for most of the energy dissipation, CPU and main memory. Recent statistics state that CPU consumes 30\%-60\% of the system power~\cite{7279063}. Several techniques to reduce that energy consumption are designed and adopted~\cite{976921, dvfs2, 1047758, 995696}. Dynamic Voltage and Frequency Scaling (DVFS)~\cite{dvfs2} and Dynamic Power Management (DPM)~\cite{1047758} are the two state-of-the-art approaches to compensate for the power and energy consumption of CPU. DPM blocks the power to the processor when it is in idle state while DVFS dynamically adjusts the clock cycles and voltages of the CPU.

On the other hand, 20\%-48\% of the energy consumption is attributed to the main memory~\cite{7279063, 1250880, 4404806}. Traditional main memory systems are composed of homogeneous memory modules, mainly DRAM which is a volatile, high bandwidth and low latency memory device. But it consumes significant energy due to volatility, destructive read operations, and refresh energy.
CPU-based energy reduction methodologies are also studied for DRAM as well, like powering down the memory ranks, controlling the base memory voltage and frequency~\cite{David:2011:MPM:1998582.1998590, 4658648}.
However, these techniques do not fulfill the performance and energy requirements per application. Whereas, using DPM and DVFS at memory level degrades the overall system performance due to state transition latency. Specifically, these approaches only enable system-level power control, which cannot meet the performance requirements of various applications. Some applications need more computation while others frequently access memory to perform read and write operations.

New materials to design memory devices, such as Spin-Transfer Torque RAM (STT-RAM), Phase Change Memory (PCM), Magnetic RAM (MRAM), and 3D-XPoint, are being studied to either use as main memory or in conjunction with traditional memory, DRAM. On the other hand, these devices do not have idle energy consumption, which makes them suitable for reducing energy consumption. These Non-Volatile Memory (NVM) devices, such as STT-RAM and 3D-XPoint, make it a more suitable alternative than DRAM as the main memory due to specific properties such as byte-addressability, persistence, high density, and less energy consumption~\cite{sttramdensity, sttram, Ramos:2011:PPH:1995896.1995911}. However, NVM offers lower bandwidth and longer latency than DRAM. Therefore, it cannot serve as a complete replacement of DRAM.  Thus, Hybrid Main Memory System (HMMS) has been proposed that incorporates both DRAM and NVM on the processor memory bus\cite{Qureshi:2009:SHP:1555754.1555760, moca, xmem, unimem, awais2019}.

\begin{table*}[!bp]
	\caption[Specs of NVMs]{\small Specification of NVM devices and normalized energy of memory command/byte in nano-Joules~\cite{3dxpoint, siena, sttram_latency, 7208275, sttram, pcm_energy}}
	\label{tab:tabl_nvms}
	\begin{center}
		\begin{tabular} {cccccccccc}
			\hline
			 Memory & BW  & Latency & Endurance & Row Read  & Row Write & Refresh \\
			 Device & ($GB/s$) & (ns) && Energy (nJ) & Energy (nJ) & Energy (nJ)\\
			\hline
			 DRAM & 25.6 & 10-50 & $10^{16}$ & 3.15 & 3.23 & 0.94 \\
			 STT-RAM & 10.6 & 32-72 & $10^{15}$ & 2.86 & 7.68 & 0 \\
			 PCM & 3.5 & 50-100 & $10^8$-$10^9$ & 1.39 & 34.55 & 0\\
			\hline
		\end{tabular}
	\end{center}
\end{table*} 

The energy consumption at the application level depends on the nature of the application workloads and the access characteristics of its memory variables. Application energy consumption varies with different workloads as the application's memory object access patterns, such as lifetime, size, accessed volume, read/write ratio, spatial \& temporal locality, and sequentiality, change with the workload~\cite{deepmap}. Further, various memory devices such as DRAM, PCM, and STT-RAM exhibit different characteristics for performance and energy consumption, as shown in Table~\ref{tab:tabl_nvms}. 
In HMMS, optimally placing memory variables to a specific memory module will lead to optimized performance and high energy efficiency~\cite{xmem}.
For example, a write-intensive variable will consume more energy at the memory module, which has high write energy, so it will be efficient to place that variable to a memory module that consumes less write energy. 
Thus, placing memory objects on NVM devices by considering their characteristics is likewise essential. 

Several works to place memory objects in HMMS have been proposed~\cite{moca,unimem,xmem}. These works classify the memory objects in an application into several categories, such as bandwidth, latency, streaming objects, and pointer tracking objects, and assign the application to the most appropriate memory module~\cite{xmem, unimem}. Memory Object Classification and Allocation framework (MOCA)~\cite{moca} optimizes the performance of the ternary HMMS by placing memory objects in the best-suited memory module. It considers their access behavior specifically based on the rate of the Last-Level Cache misses per kilo instruction (LLC MPKI) and reduce the energy consumption through object placement. The major goal of MOCA is to improve the performance of HMMS by selectively placing memory objects meanwhile, through placement, it reduces the energy consumption as well. However, only considering the LLC MPKI does not provide an optimal placement of memory objects in HMMS to optimize the performance and energy efficiency. As other access behaviors of the memory objects, such as lifetime, size, and accessed-bytes, play an important role in the performance and energy consumption of the application. MOCA only provides a static placement and does not consider the energy consumption requirement during the application execution time.

\begin{figure*}[htb!]
	\centering
{\begin{tabular}{ccc}
	\centering
			\includegraphics[width=0.4\textwidth]{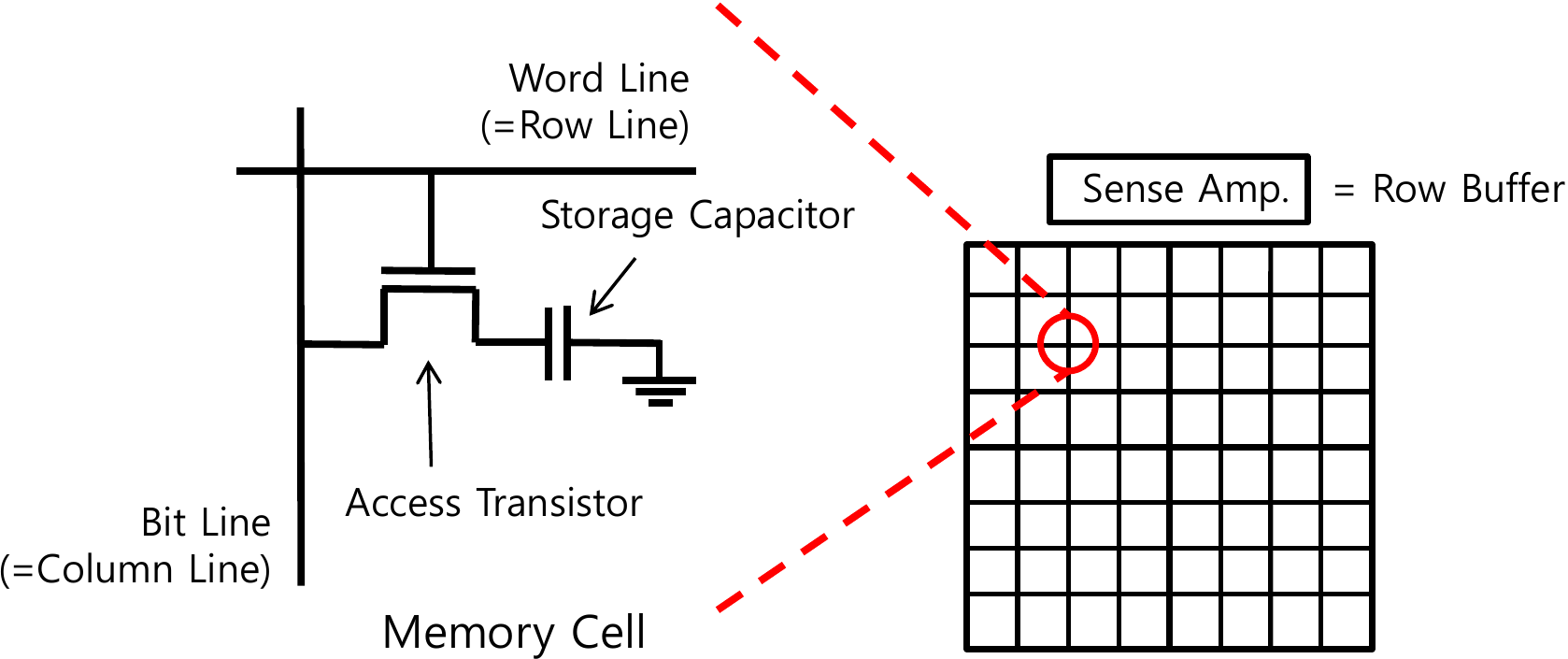} &
			\includegraphics[width=0.4\textwidth]{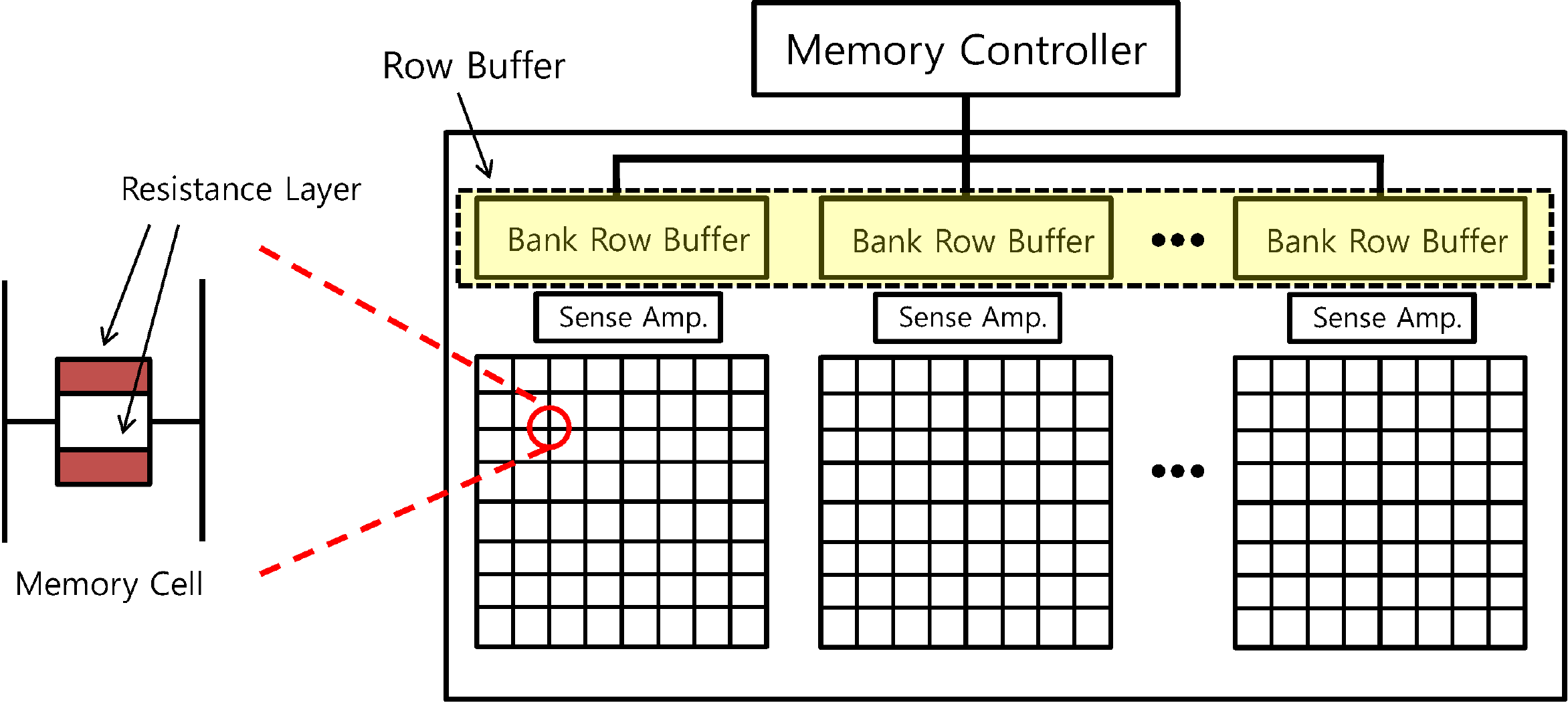}  \\
	 {\small (a) Structure of DRAM array}&
	 {\small (b) Structure of STT-RAM} &\\
\end{tabular}}
\caption{Comparison of DRAM and STT-RAM cell architectures} 
\label{fig:architecture}
\vspace{-0.2in}
\end{figure*}

In this paper, we propose \emap{}, which is an optimal memory object placement algorithm based on object level profiling information and ILP-based placement algorithm. \emap{} considers the fine-grained memory objects access patterns and per-object energy consumption of an application to provide optimal placement policies for memory objects to meet the energy limiting constraint while optimizing performance in HMMS.  \emap{} is equipped with two placement modules, \emplan{} and \emdyn{}. The \emplan{} is a static module that determines static placements of objects before applications begin to run. It optimizes the application performance while reducing the energy consumption to a specific rate by optimally placing memory objects in HMMS. The \emdyn{} is a dynamic module that reduces the energy consumption while optimizing an application's performance by re-evaluating the object placement and migrating those objects if necessary to satisfy the energy requirement during application runtime. 

This paper provides following specific contributions: 
\squishlist
\item
\emplan{} employs the memory object profiler, Integer Linear Programming (ILP) based Energy Estimator, Placement Planner, and a Runtime Memory Allocator. 
In \emplan{}, the memory profiler analyzes the diverse access patterns of memory objects of applications using a Two-Pass memory profiler~\cite{deepmap}.
The Energy Estimator considers the energy consumption and characteristics of both devices, DRAM and NVM, in HMMS respectively. 
The Placement Planner calculates the optimal placement of memory objects by considering the object access patterns and the energy consumption obtained from the Energy Estimator.
The Runtime Memory Allocator allocates memory objects to respective memory modules according to the placement policies obtained by Placement Planner. 

\item
\emdyn{} consists of an ILP-based Migration Planner and Migration Executor.
While \emplan{} decides the placement of objects to optimize the performance and meet the given energy limiting constraints, the runtime memory allocator of \emdyn{} can re-allocate the objects following the decided placement during the application execution. 
\emdyn{} changes the optimal placement decision at runtime of the application as the energy constraint or the request to reduce the more energy consumption can be placed by the user or system. 
\emdyn{} considers the incoming energy change request and anticipates the memory objects migration by considering their access patterns and obtain a new placement.
\emdyn{} only migrates those memory objects which are already allocated while newly allocated objects are now placed by considering new optimal placements.

\item
We evaluate the proposed \emap{} using real-time application benchmarks, such as NAS Parallel Benchmark (NPB)~\cite{Bailey2011} and Problem Based Benchmark Suite (PBBS)~\cite{pbbs}. We evaluate the proposed \emap{} on two different testbed configurations. Testbed I is an IBM server, whereas, Testbed II is an Intel based Non-Uniform Memory Access (NUMA) servers. Due to the lack of actual device, we emulated STT-RAM over DRAM using the emulation platform, QUARTZ\cite{quartz}. We compared our solution with MOCA framework~\cite{moca}. The evaluation results show that our \emplan{} outperforms MOCA in reducing the energy consumption up to 14\% with the same performance on both of the application benchmarks. Our proposed \emdyn{} also meets the performance and energy requirements for NAS benchmark with negligible migration cost in terms of migration time and energy consumption. The average energy efficiency of \emdyn{} is up to 4\% with considering migration cost. 

\squishend

\section{Background} 
\label{sec:background}
This section provides the background on the object placement in HMMS, our candidate NVM device, and the object profiling.

\subsection{Spin-Transfer Torque RAM (STT-RAM)} \label{sec:stt-ram-arch}
STT-RAM is one of the rapidly developing memory technology. The characteristics of STT-RAM as shown in Table~\ref{tab:tabl_nvms} categorize it as one of the most suitable candidates for this work as it has low latency and high durability. However, one disadvantage of STT-RAM is that it has a high write energy consumption. A recent study states that STT-RAM has more than twice of energy consumption in writing to a memory array than DRAM~\cite{hameed}. 
To deal with this, we adopted the partial write methodology as one of the energy optimization methods of STT-RAM~\cite{sttram}.

Figure~\ref{fig:architecture} shows the comparison of the memory cell structure of DRAM and STT-RAM. As shown in Figure~\ref{fig:architecture}(a), DRAM memory cell stores data in a storage capacitor. When a memory row is read, charge sharing occurs between the precharged bit line and the storage capacitor.
This destroys the data stored in the cell. Due to this destructive read, DRAM must perform a restore operation, which requires the sense amplifier to re-write the sensed data to the memory cell. Therefore, sense amplifier should maintain the data in itself, which acts as a row buffer in DRAM. However, since STT-RAM performs non-destructive reads, its row buffer and sense amplifier exist separately and act independently from each other as shown in Figure~\ref{fig:architecture}(b). Thus, when STT-RAM array write occurs, updates are first made to the row buffer. If memory access whose address is not fetched to the row buffer, a row buffer conflict occurs. The row buffer write-back is operated and effective memory array write is done.

In addition, this mechanism incurs unnecessary energy consumption. The ~\cite{sttram} states that when an STT-RAM row buffer conflict occurs, the data in row buffer is clean more than 60\% and it is less than 6\% that the number of dirty cache blocks in a row buffer is more than four. That is, a large portion of row buffer is unmodified at row buffer conflict, and if it was modified, the number of modified blocks in row buffer is generally less than 4 cache blocks. But, without any optimization, the whole row buffer should be written back though most of the blocks are clean and it incurs severe energy consumption in STT-RAM. To mitigate this problem, \cite{sttram} proposed an optimization method, partial-write, which writes back only the dirty blocks when a row buffer conflict occurs by holding dirty bits of all cache blocks of row buffer in the memory controller. When the row buffer is 4 KB, only 64 bits of space is required. Therefore, it is spatially feasible and the energy consumption of the STT-RAM can be reduced by upto 70\%.

In this work, we target STT-RAM specifically as the energy model of STT-RAM is already presented, while for other NVM devices, there is no any energy model. Besides, the adoption of our work on PCM is part of the future work as it requires considering some architectural choices as for STT-RAM, we have to consider the Row Buffer. There is no such architectural design presented for PCM yet. 

\subsection{Object Placement in HMMS}
Memory devices such as High Bandwidth Memory (HBM), Reduced Latency DRAM (RLDRAM), and Low Power DDR (LPDDR) are being produced and studied in a considerable pace~\cite{siena}. On the other hand, PCM and STT-RAM are the two most rapidly growing NVM devices to be placed on a processor-memory bus in conjunction with DRAM to enable HMMS~\cite{pittir20187, moca, xmem, unimem, memif, rthms, task-para}. Various works~\cite{moca, xmem, unimem} have already been studied to place memory objects on different memory modules in HMMS by considering their characteristics.

Nevertheless, one type of memory cannot satisfy various demands at the same time as these memory devices have different read/write access latency, density, and energy utilization. For example, RLDRAM has low latency whereas, power and energy consumption are five times higher than the DRAM. 3D-XPoint has 750 times higher density than DRAM, but the latency is 1,000 times higher than DRAM~\cite{moca}. If the major workload in the system requires both low latency and high density, then the main memory configuration with either RLDRAM or 3D-XPoint will not produce optimal results. However, if the main memory is configured by using those two types of memory together, it can achieve optimal results by placing latency-sensitive objects in RLDRAM and large-sized objects in 3D-XPoint. Therefore, several studies have shown interest in enabling performance efficient options to allocate memory objects in HMMS~\cite{moca, xmem,6267902, unimem}. 
Our target memory system is HMMS environment comprised of DRAM and NVM, where DRAM has high performance and NVM has high density and power-efficiency. 

In addition, usually the different applications exhibit varying characteristics according to their object-level access patterns. In HMMS, placing memory objects according to their object-level information can be helpful in optimizing the performance and reducing the energy efficiency~\cite{xmem}. For example, scientific applications majorly work on their dynamically allocated memory objects and different objects exhibit different properties~\cite{deepmap}. Now if an application allocate two objects, i.e., A and B. If object A is read-intensive which means that application will mostly read that object while object B is write-intensive. Now placing both objects in homogeneous memory system will lead to high energy consumption due to object B. On the contrary, in HMMS, placing these objects will be non-trivial as if the read-intensive object will be placed in high latency memory module than it will degrade the performance and if the write-intensive object will be placed in energy sensitive device than it will consume high amount of energy. So, in HMMS, optimally placing these objects being aware of access patterns and device properties will lead to optimal performance and high energy efficiency. 

\subsection{Object-level Memory Profiling}
As the memory device type has an effect on energy consumption, the memory object access patterns also play a vital role in energy dissipation. For example, a write-intensive object will consume more energy in a memory device with high write energy. Therefore, the placement of objects on the basis of the object access pattern and NVM device characteristics will lead to optimized performance and energy efficiency. In this paper, we adopted the two-pass memory profiler to extract object access patterns information~\cite{deepmap}. We utilize that extracted information to estimate the energy consumption for objects to be placed on HMMS, and the device specific energy model is explained in the Section~\ref{design}. 

\begin{figure}[t!]
	\centering
	\includegraphics[width=0.4\textwidth]{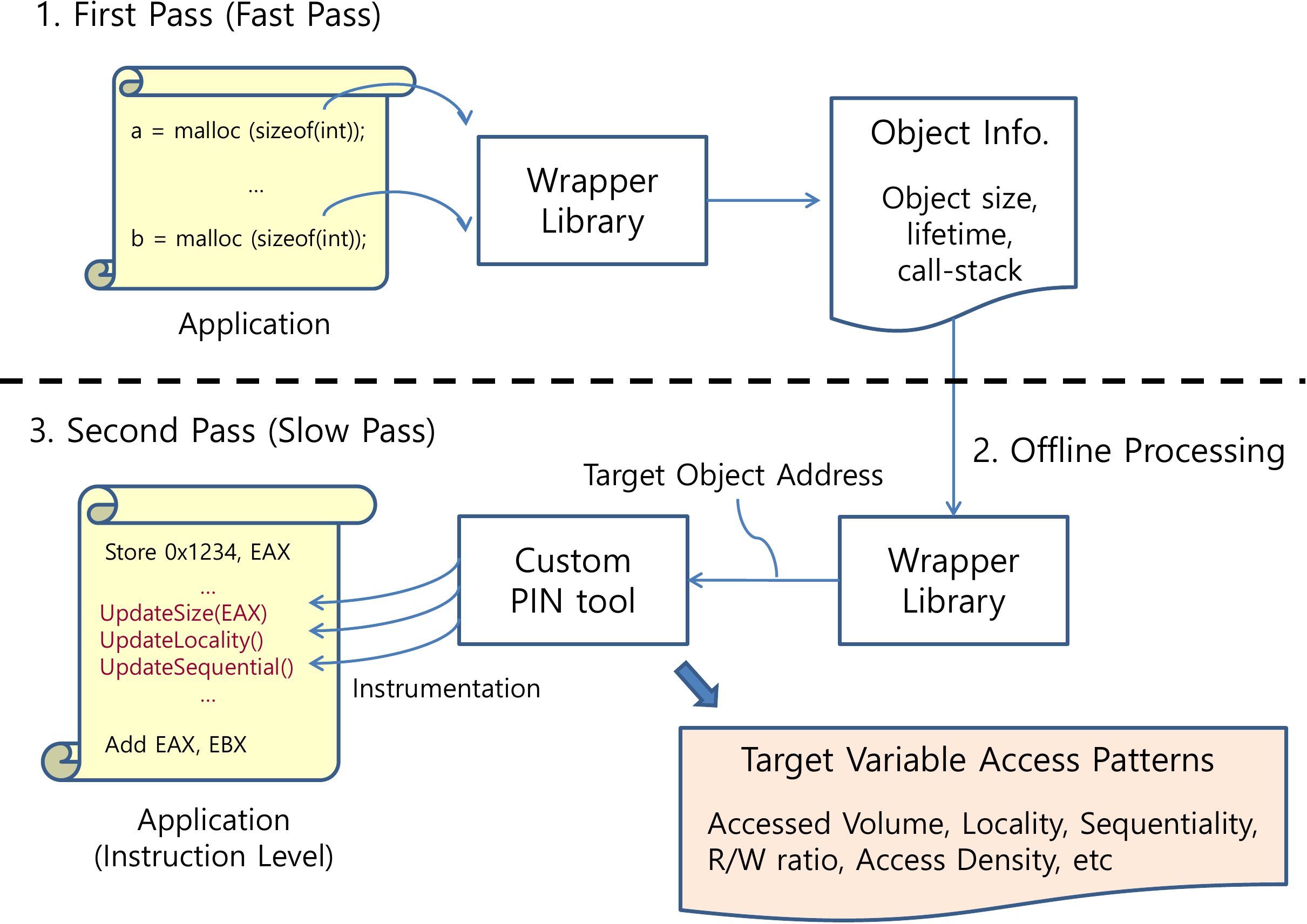} 
	\caption{Two-pass Memory Profiler~\cite{deepmap}}
	\label{fig_profiler}
	\vspace{-0.2in}
\end{figure}

Two-pass memory profiler targets the dynamically allocated heap memory variables and extracts basic information such as size, lifetime, and call-stack. We have extended the two-pass profiler to extract the fine-grained object access patterns and the details are provided in section~\ref{object-profiler-extended}. The distinction between variables and objects is based on a call-stack consisting of the order of memory allocation function calls and the order of allocation function return addresses. The two-pass profiler workflow is shown in Figure~\ref{fig_profiler}.

\section{Heap Memory Object Placement System}
This section describes the design goals, various components, and the interactions between components in the \emap{} system.
\vspace{-0.1in}
\subsection{Goals}
In this section, we discuss our key design principles.

{\textit{\textbf{Optimal Object Placement:}}} The high access latency of NVM devices makes them ill-suited to replace the main memory. Using NVMs in conjunction with DRAM forms a HMMS. It helps in reducing the high access latency of NVM through intelligently placing memory objects in DRAM and NVM. This first goal of \emap{} is to obtain the optimal placement for heap memory objects by considering their detailed access patterns, such as lifetime, size, accessed volume, and dirty cache-lines, for an HMMS. 

\textit{\textbf{Energy Efficiency:}} The idle energy consumption of DRAM makes it a power-hungry device. On the other hand, NVMs do not have idle energy utilization, which makes them a suitable candidate for reducing the energy efficiency of the system. But, the dynamic energy consumption of the NVM devices is high, specifically when writing. So, placing memory objects in HMMS effectively will help in reducing the energy efficiency of the system. The Second goal of \emap{} is to optimize the energy efficiency of the HMMS by optimally placing the memory objects. 

To achieve these goals, we proposed \emap{}, the methodology to place the memory objects in the HMMS by considering their detailed access patterns. In particular, we developed an Integer-Linear Programming based memory object placement planner to efficiently allocate the memory objects in the HMMS while meeting the energy requirements of the system.

\begin{figure*}[t!]
	\centering
	\includegraphics[width=0.7\textwidth]{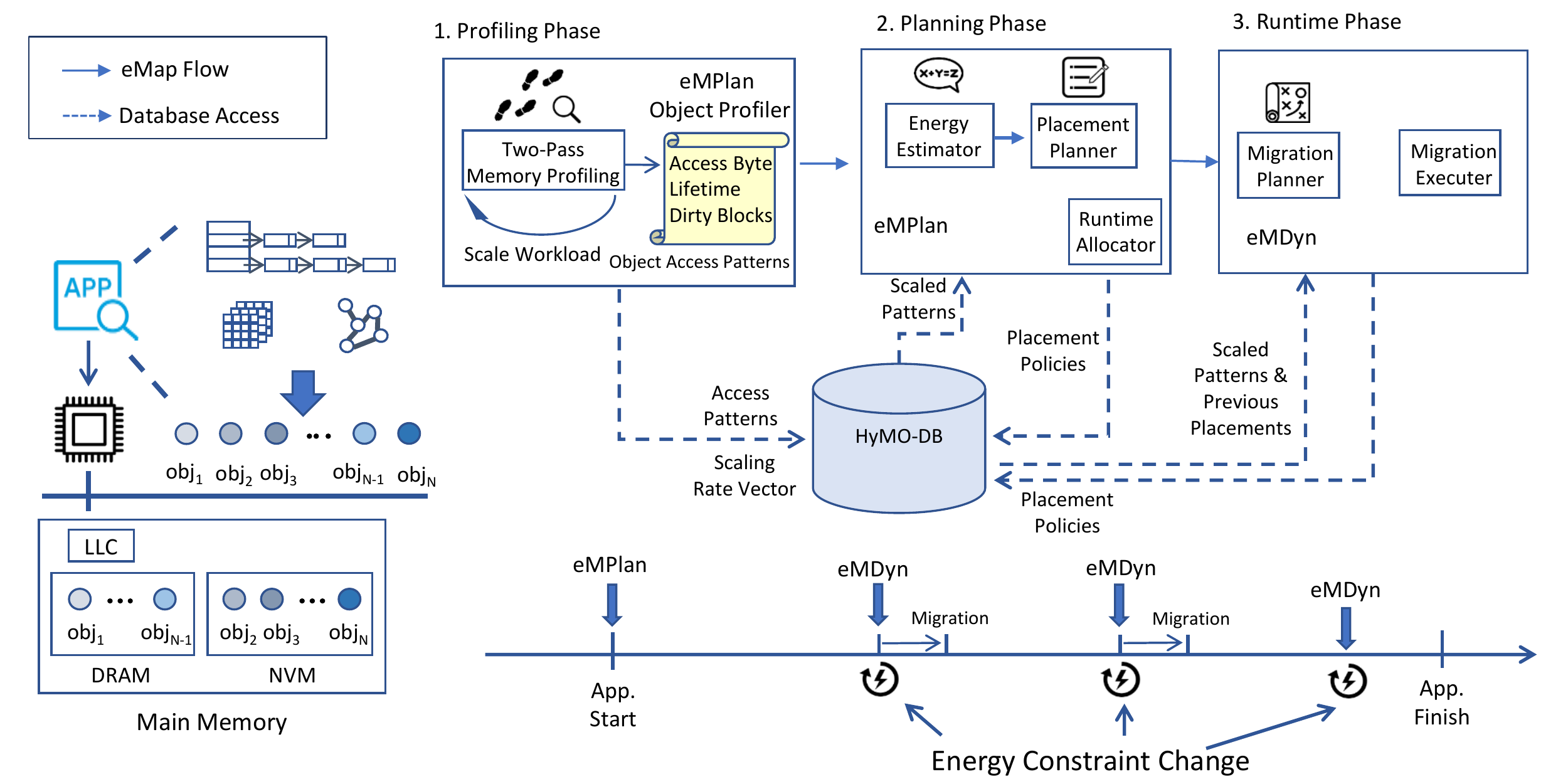}
	\caption{Description of various components of \emap{} and how they interact} 
	\label{fig_overview}
\end{figure*}

\subsection{Overview}
Figure~\ref{fig_overview} depicts the interaction between various components of \emap{} for HMMS, composed of DRAM and STT-RAM. The left side of the diagram shows the execution of an application in HMMS where it allocates some of the memory objects in DRAM while some in STT-RAM. The right side of the figure shows three phases for \emap{}, Profiling, Planning, and Runtime. The profiling and planning phases are the part of our static module of \emap{}, while the runtime phase belongs to the dynamic module. 

\textbf{\textit{Profiling Phase:}}
It adopted Two-Pass memory profiler for the extraction of memory-level object access patterns such as size, lifetime, accessed volume, and last-level cache (LLC) miss counts~\cite{deepmap}. These extracted object access patterns are stored in the database named Hybrid Memory Object Database (HyMO-DB), as shown in Figure~\ref{fig_overview}. HyMO-DB also stores the device-level characteristics and the placement decisions of the memory objects from planning and runtime phases. 

\textbf{\textit{Planning Phase:}}
It is an ILP-based algorithm and employs three major constraints, i.e., (i) Decision, (ii) Capacity, and (iii) Energy. The pseudo-code of planning phase is shown in Algorithm~\ref{algorithm}. The object access patterns for an application are fetched from HyMO-DB and the placement decisions are generated. 
\begin{itemize}
    \item In the first step (lines 1 to 4), the ILP model is loaded using a third party library~\cite{lpsolve} and the Decision constraint is defined for each memory object from the HyMO-DB of an application. Decision constraint is bound to be binary (either 0 or 1) as our target HMMS consists of 2 memory devices, DRAM and STT-RAM. 
    \item In the second step (lines 5 to 9), the Capacity constraint is defined for all the memory objects. 
    Capacity constraint is bound not to exceed the memory modules' capacity for the placement of memory objects, which means that the number of objects that are placed on each memory module should not exceed the memory device's capacity individually. 
    \item In the third step (lines 10 to 13), the Energy constraint is defined to reduce energy efficiency. As one of the primary goals of our proposed algorithm is to reduce the energy efficiency by optimally placing the memory objects in HMMS by considering the energy requirement, we take the rate of energy consumption to be reduced from DRAM energy consumption as input and bound the ILP constraint to not exceed for each memory object.
    \item 
    In the fourth step (lines 14 to 16), we define the objective function of our proposed algorithm, which is to optimize the performance, determine the overall latency of each memory object, and bind the objective function according to the latency values. 
    \item 
    Last but not the least step (lines 17 to 19), we optimize our ILP algorithm for minimum values so that the performance is optimized. Then we write the ILP model with all the above three constraints and compute the model for the optimal placement decisions. Once the placement is calculated, it is stored in the HyMO-DB, and the application is executed with static placements. 
\end{itemize}  

\textbf{\textit{Runtime Phase:}}
\emdyn{} plays a vital role during the execution of the application, as there may be a need to change the energy limiting constraint during the application execution. In the Runtime phase, the migration planner is triggered which re-evaluates the placement of memory objects by considering their current states, such as where an object is placed and how much lifetime of the object is remaining, and obtains a new placement policy for all the major objects. The migration planner is also based on the ILP algorithm shown in Algorithm 1 with just modifications in the computation part of the energy consumption. Once the new placement is obtained, the object is either migrated based on the placement decision from its previously placed memory module to the new memory module that is from DRAM to NVM and vice versa by \emdyn{} migration executor module. 

\begin{algorithm}
\DontPrintSemicolon
  
  \KwInput{Object access patterns and energy rate}
  \KwOutput{Placement decisions}

    Load ILP model\;
    \tcp{\small{Decision Constraint}}
   \While{objects}
   {
   		Add Constraint\;
   		Bound to be binary\;
   }
   
   \tcp{\small{Capacity Constraint}}
   \While{Objects}
   {
        set objects.size $\rightarrow$ ILP Format\;
   }
   Add Constraint\;
   Bound constraint $\leq$ DRAM capacity\;
   Bound constraint $\leq$ NVM Capacity\; 
   
   \tcp{\small{Energy Constraint}}
   \While{Objects}
   {
        set object.energy $\rightarrow$ ILP Format\;
   }
   Add Constraint\;
   Bound constraint $\leq$ energy.DRAM * Energy Rate\;
   
   \tcp{Objective Function}
   \While{Objects}
   {
        set object.latency $\rightarrow$ ILP Format\;
   }
   Load ILP\_Objective\_Function\;
   
   \tcp{\small{Compute Model}}
   set\_minim(ILP\_model) \tcp*{\footnotesize{Optimize for minimization}}
   write\_ILP(ILP\_model)   \tcp*{\footnotesize{Write the model with all the constraints}}
   sovel\_ILP(ILP\_model)   \tcp*{\footnotesize{Solve the model}}
   return Object\_placement   \tcp*{\footnotesize{Return object placement to HyMO\-DB}}
\caption{ILP-based object placement algorithm }
\label{algorithm}
\end{algorithm}

\section{Design and Implementation} \label{design}
In this section, we explain the details of our proposed \emap{} approach.

\subsection{\lowercase{e}Mplan: Static Object Placement}\label{sec:emplan}
This section provides design details of the \emplan{}.

\subsubsection{Object Profiler} \label{object-profiler-extended}
\emplan{} profiles the memory object access patterns and estimates the energy consumption of the object with device specific energy model and extracted object access patterns. We extended the Two-Pass memory profiler~\cite{deepmap} to extract fine-grained profiling information of heap memory objects. As shown in Figure~\ref{fig_profiler}, Two-Pass Profiler operates in two passes, i.e., Fast-pass and Slow-pass. Fast-pass identifies all the heap memory allocations using the call-stack and assign a hashed identifier and the size of the objects are also obtained. In the offline processing (when application is not being executed), target memory objects are selected for detailed profiling. For effectiveness and to reduce the complexity of profiling, we only take into account those memory objects whose accessed size is larger than 1 MB, called major objects\footnote{Our solution only considers the major objects for the placement and we interchangeably used the terms major object, heap memory objects and simply the objects.}. 

The Slow-pass then considers the target objects selected in the offline processing and extracts the detailed access patterns. The Slow-pass utilized customized PIN-Tool~\cite{pintool} which can easily be extended to extract all the necessary object-level access patterns at instruction-level. Two-Pass profiler provides a wrapper library for tracing the heap memory allocation calls, such as malloc, realloc, and calloc, and each heap memory object access goes through the custom analysis code which is based on PIN-tool. We extended the wrapper library to extract the required object access information using the PIN-Tool. We traced all store instructions to the heap-allocated objects and calculated the various object access patterns. Also, we used the Performance API (PAPI)~\cite{papi}, a hardware event counter, in the custom analysis code to count the cache misses. 

For STT-RAM row buffer, we set a buffer to have the same size in the virtual memory of profiler. This virtual buffer is used to count the number of dirty cache blocks which will be written back to the memory array when a row buffer conflict occurs. The assumption here is that when a single process runs on the machine, the number of dirty cache blocks in the virtual memory buffer and the actual device's row buffer will exhibit a consistent pattern if the sizes of both buffers are identical. 

For DRAM page policy, we take into account the closed page policy, which always flushes open row buffer to corresponding row in the memory array. While open page policy has different types, for example fixed open page policy and adaptive open page policy, we did not consider open page policy concepts because we just deal with the object placement in HMMS, not optimization of DRAM memory controller. Thus, we assumed general memory controller page policy.

The remaining memory objects that are less than 1 MB are placed in DRAM. And the baseline placement of memory objects and the code of the application is DRAM. To estimate the energy consumed by the objects in HMMS, the following memory access information is required: (i) object size, (ii) total amount of memory read and written by object, accessed volume (iii) object lifetime, and (iv) the total number of dirty cache blocks in a certain size of row buffer.

\subsubsection{Scaling Rate Vector} \label{scaling}
When an application's workload changes its access patterns are also varied accordingly. However, \cite{deepmap} states that 98.1\% of the objects are scaled or fixed as the input workload size scales. This means that when input workload scales, object access patterns also scale consistently (with a scaling rate of 1 for a fixed object). Thus, the profiling of target application is not required for every time the workload changes and a scaling rate vector can be derived. 
The scaling rate vector of the access patterns is based on the profiling information of various workloads of the application so it can be stored in the HyMO-DB shown in Figure~\ref{fig_overview}.

The input size of the application can be set by user which makes the derivation of scaling rate vector easy. For example, if the workloads of the application are 'N', we can derive the average rates of access patterns among the 'N' input sizes and compose the vector with these rates. A generalized view of the scaling rate vector for various access patterns can be shown in Equation \ref{eq:scaling} where $ap_i$ is a particular access pattern, such as size, lifetime and LLC miss count, $in_i$ is the workload size of the target application, and N is the total number of workload a target application provides. 
\begingroup\makeatletter\def\f@size{7.5}\check@mathfonts
\begin{equation}\label{eq:scaling}
	avg\_grad = \sum_{i=1}^{N-1}{\{(ap_{i+1} - ap_{i})/(in_{i+1} - in_{i})\}}/(N-1)
\end{equation}

For example, if the $i$-th object has size $(S_i)$ is 10 MB for the workload $in_1$, $S_i=19 MB$ for the workload $in_2$ and $S_i=25 MB$ for the workload $in_3$ then the scaling rate vector for the size of $i$-th object can be derived as: 
\begingroup\makeatletter\def\f@size{7.5}\check@mathfonts
\begin{align*}
         {\{(19 MB - 10 MB)/(in_2 - in_1)}{+ (25 MB - 19 MB)/(in_3 - in_2)}\}/2
\end{align*}

\subsubsection{Energy Estimator}
The energy estimation in \emplan{} is the key component as it provide the estimated energy consumption to compute the optimal placement of memory objects. The energy estimator calculates the per-object energy consumption for both of the memory modules of HMMS where all the objects are placed to either of the memory devices, respectively. We adopted STT-RAM as an example of NVM device and suggest the energy model of DRAM and STT-RAM based on the methodology~\cite{sttram}. In this work, we target STT-RAM specifically as the energy model and the architectural details are provided in \cite{sttram} while other NVM devices architectural details and the energy models are not yet determined. 

The memory commands are classified as Activate (ACT), Pre-charege (PRE), Read/Write (RD/WR), Refresh (REF), Row Buffer Access (RBA), and Write-Back (WB)~\cite{sttram}. 
ACT is the command which activates the accessed bank and row before memory RD/WR in both memory devices. 
PRE pre-charges the bit-line to prepare the next memory access and to restore the read or written data in memory array of DRAM. 
RD/WR are the actual memory read and write. 
REF recharges the voltage to storage capacitor of memory cell to prevent a data loss due to current leakage in DRAM. 
RBA is the cost to access the row buffer and WB is the cost of writing row buffer data back to memory array when a row buffer conflict occurs in STT-RAM.
Table~\ref{tab:table_energy_norm} shows the per-byte energy consumed by above mentioned commands.
\begin{table}[!b]
	\caption[Energy Consumption of Memory Command per Byte in nano-Joules]{Energy consumption of memory command per byte~\cite{sttram}}
	\label{tab:table_energy_norm}
	\begin{center}
		\begin{tabular} {lllllllllll}
		    \hline
			 Memory Command & Energy (nJ) \\
			 DRAM Activate+Pre-charge & 3.07 \\
			 DRAM Read/Write & 1.19 \\
			 DRAM Refresh & 0.35  \\
			 STT-RAM Activate+Pre-charge & 2.68  \\
			 STT-RAM Row Buffer Access & 1.00  \\
			 STT-RAM Write-Back & 2.83  \\
			\hline
		\end{tabular}
	\end{center}
\end{table} 

Our proposed energy model calculates the energy consumption on per-object basis.  Equation~\ref{eq:dram} represents the energy consumption of the $i$-th object when it is placed in DRAM. The accessed volume ${(AV_i)}$ of the object represents how much in total an object is being accessed during its lifetime which is extracted during the profiling phase. It is reasonable to multiply ${(AV_i)}$ with DRAM ACT, PRE and REF energy consumption. For DRAM refresh energy ($dE_{REF}$), we assumed selective refresh policy per row (4 KB). In addition, the refresh energy of DRAM is also being considered with the lifetime ${(T_i)}$ and actual size ${(S_i)}$ of the object. The reason to consider the accessed volume and size separately is to comprehensively take into account all the read and write operations that are being performed for object during its lifetime.
Equation~\ref{eq:sttram} represents the energy consumption of the $i$-th object when it is placed in STT-RAM. As shown in the section~\ref{sec:stt-ram-arch}, STT-RAM read/write operations fall back to Row Buffer that's why we have considered the row buffers exclusively while considering the read/write operations to STT-RAM. Same as DRAM, STT-RAM also bears the cost of ACT and PRE. In addition, we have considered the write backs to STT-RAM in terms of number of dirty cache blocks $(N_{DC})$ and the cache block size $(V_{CB})$. Table~\ref{tab:equation_notations} defines the notations used in the equations. 
\begingroup\makeatletter\def\f@size{7.5}\check@mathfonts
\begin{equation} \label{eq:dram}
   {DE_{i} = dE_{A+P}\cdot AV_i + dE_{RW}\cdot AV_i + dE_{REF}\cdot S_i\cdot T_i}
\end{equation}
\begingroup\makeatletter\def\f@size{7.5}\check@mathfonts
\begin{equation}\label{eq:sttram}
   {NE_{i} = nE_{A+P}\cdot AV_i + nE_{RBA}\cdot AV_i + nE_{WB}\cdot N_{DC}\cdot V_{CB}}
\end{equation}

\subsubsection{Placement Planner}
The Placement Planner of the \emplan{} determines the optimal placement of memory objects to optimize the performance while satisfying the energy limiting constraint that is requested externally. It utilizes the per-object energy estimation model for DRAM and STT-RAM. We modeled the Integer-Linear Programming (ILP) algorithm for the Placement Planner. Our model is based on three major constraints for the implementation of Placement Planner, Decision Constraint, Capacity Constraint, and Energy Limiting Constraint. We adopt a third-party shared library, lp\_solve~\cite{lpsolve}, to implement these constraints.
\paragraph{\textit{Decision Constraint}} 
The Decision Constraint is to make the placement decision for each memory object that whether a particular object will be placed on DRAM or NVM. This placement can be represented by an ILP variable, ${X_i}$, which represents 0 for NVM and 1 for DRAM as shown in equation~\ref{eq_decision}. 
\begingroup\makeatletter\def\f@size{7.5}\check@mathfonts
\begin{equation} \label{eq_decision}
	0 \le X_i \le 1 \quad \mbox{for } i=1,2,...,N
\end{equation}

\paragraph{\textit{Capacity Constraint}} \label{sec:capacity-emplan}
The second constraint takes into account the limited capacities of memory devices. It checks that all the allocated objects sizes should not exceed the capacity of memory device. Equation~\ref{eq_capacity} shows the capacity constraint for both memory devices. $C_{D}$ represents DRAM while $C_{N}$ is NVM capacity.
\begingroup\makeatletter\def\f@size{7.5}\check@mathfonts
\begin{align} \label{eq_capacity}
    \sum_{i=1}^{N}{X_i\cdot S_i} \le C_{D} && \sum_{i=1}^{N}{(1-X_i)\cdot S_i} \le C_{N}
\end{align}

\paragraph{\textit{Energy Constraint}}
The third constraint considers the energy limitation requests issued by client or the remaining battery lifetime of the system. The external energy limit constraint is given as a specific ratio of existing energy consumption. All objects of target application must be allocated not to exceed the required ratio of the energy which is consumed when all objects are placed in DRAM. Equation~\ref{eq_energy} shows the energy limiting constraint. Let the required ratio be ${R}$, then the sum of energy consumption of all the objects placed in HMMS should not exceed ${R}$ times the energy consumption of objects placed entirely in DRAM (${DE_i}$). 
\begingroup\makeatletter\def\f@size{7.5}\check@mathfonts
\begin{equation} \label{eq_energy}
	\sum_{i=1}^{N}{\{X_i\cdot DE_i + (1-X_i)\cdot NE_i\}} \le \sum_{i=1}^{N}{DE_i}\cdot R
\end{equation}

\paragraph{\textit{Objective Function}}
The goal of \emplan{} is to minimize memory access latency while satisfying the above constraints. That is, the sum of whole HMMS access time should be minimized.
Each device access time can be derived by multiplying the total access counts of objects and the latency of the device. The Performance API~\cite{papi} is used to count the actual memory access in profiling step to get the total LLC miss counts. This objective is presented in Equation~\ref{eq:objective} where $L_{DRAM}$ and $L_{NVM}$ indicate the latency of DRAM and NVM respectively, and $L3M_i$ indicates the LLC miss count of $i$-th object. 
\begingroup\makeatletter\def\f@size{7.5}\check@mathfonts
\begin{equation}\label{eq:objective}
	f = \sum_{i=1}^{N}{\{X_i\cdot L_{DRAM}\cdot L3M_i + (1-X_i)\cdot L_{NVM}\cdot L3M_i\}}
\end{equation}

\subsubsection{\textit{Runtime Allocator}}
Once the Placement Planner decides the placements for all the major variables, the target application is executed in real-time and the Runtime Allocator of \emplan{} operates to allocate those objects. The Runtime Allocator configures the object allocation table with the determined placement at the initialization step. In the object allocation table, the identification of objects is achieved with the hash values of the call-stack of dynamic allocation functions. Once the target application starts execution, \emplan{} hooks all the dynamic memory allocation functions on every object and calculates the hash value from its call-stack and compares with the object allocation table to identify the target objects. If the allocated object is the placement target object, the placement decision of the object is referred from the object allocation table. If the mapped device is DRAM, existing allocation functions such as malloc is used. 
If the allocated device is NVM then NVM allocation API, which is provided by NVM emulation tool QUARTZ~\cite{quartz}, is used.
\begin{table}[!b]
	\caption[Notations used in equations]{Notations used in the equations where $i$ represents the \textit{i}th object.}
	\label{tab:equation_notations}
	\resizebox{\columnwidth}{!}
		{\begin{tabular} {lllllllllll}
			\hline
			& \textbf{Notation} & \textbf{Description} &\\
			\hline
			& $dE_{A+P}$ & DRAM activate+pre-charge &\\ \hline
		    & $dE_{RW}$ & DRAM read/write &\\ \hline
			& $dE_{REF}$ & DRAM refresh & \\ \hline
			& $nE_{A+P}$ & STT-RAM activate+pre-charge & \\ \hline
			& $nE_{RBA}$ & STT-RAM row buffer access & \\ \hline
			& $nE_{WB}$ & STT-RAM write-back & \\ \hline
			& $DE_i$ & DRAM energy consumption &\\ \hline
			& $NE_i$ & NVM energy consumption &\\ \hline
			& $CP_i$ & Previous placement policy &\\ \hline
			& $dnE_i$ & Migration energy from DRAM to NVM &\\ \hline
			& $ndE_i$ & Migration energy from NVM to DRAM &\\ \hline
			& $MigCE1_i$ & Migration energy cost from DRAM to NVM &\\ \hline
			& $MigCE2_i$ & Migration energy cost from NVM to DRAM &\\ \hline
			& $T_i$ & Lifetime &\\ \hline
			& $sT_i$ & Allocation time &\\ \hline
			& $fT_i$ & De-allocation time &\\ \hline
			& $MigCT_i$ & Total migration time cost &\\ \hline
			& $dnL_i$ & Total latency from DRAM to NVM migration &\\ \hline
			& $ndL_i$ & Total latency from NVM to DRAM migration &\\ \hline
			& $MigTD_i$ & Migration time from DRAM to NVM\\ \hline
			& $MigTN_i$ & Migrate time from NVM to DRAM&\\
			\hline
		\end{tabular}} 
\end{table}
\subsection{\lowercase{e}mdyn: Dynamic object placement}\label{sec:emdyn}
\emdyn{} is the second module of \emap{} and it considers the energy limiting requests at the runtime and re-evaluate the placement of memory objects and migrate then to meet the new energy constraint. \emdyn{} is based on two sub-modules, migration planner and migration executor.

\subsubsection{Migration Planner}
The migration planner is an ILP-based algorithm to re-calculate the placement of major objects to meet the new energy requirements. Shuffling the memory objects to meet the energy constraint also incurs some energy consumption of migration, i.e., migration cost. So, it considers the access patterns, migration costs in terms of energy and performance, and the new energy limiting constraint to satisfy the energy while optimizing the performance of application in HMMS. It is also based on similar three major constraints, Migration Decision, Capacity, and Energy Constraint.

\paragraph{\textit{Decision Constraint}}
The migration decision ($X_i$) shows that if it is beneficial to migrate an object from its current placement to new one. It is similar to Equation~\ref{eq_decision}. If it is beneficial to migrate than $X_i$ will be 1 otherwise 0. 

\paragraph{\textit{Capacity Constraint}}
Similar to section~\ref{sec:capacity-emplan}, it considers that the migrated objects size should not exceed the capacity if memory devices. Let $CP_i$ be the previous placement of the object before energy constraint change. Due to space limitation, we have omitted the equations of Migration Decision and Capacity Constraint as they are equivalent to \emplan{}. 

\paragraph{\textit{Energy Constraint}}
The major goal of \emdyn{} is to meet the new energy limiting constraint while optimizing the performance. For that migration planner calculates the total energy consumption including the migration cost and then decide the new placement. The energy consumed by the objects that are being migrating from DRAM to NVM and NVM to DRAM are shown in Equation~\ref{eq_energy_dton_total} and Equation~\ref{eq_energy_ntod_total}, respectively.
\begingroup\makeatletter\def\f@size{7.5}\check@mathfonts
\begin{equation} \label{eq_energy_dton_total}
		dnE_i = DE_i\cdot \frac{t-sT_i}{T_i} + MigCE1_i + NE_i\cdot \frac{fT_i-t}{T_i}
\end{equation}
\begingroup\makeatletter\def\f@size{7.5}\check@mathfonts
\begin{equation} \label{eq_energy_ntod_total}
    \begin{split}
		ndE_i=NE_i\cdot \frac{t-sT_i}{T_i}+MigCE2_i+DE_i\cdot \frac{fT_i-t}{T_i}
	\end{split}
\end{equation}

Here, $t$ indicates the time point during the application execution when the request of energy constraint change occurred. In addition, the migration cost for energy consumption ($MigCE1_i$ \& $MigCE2_i$) for DRAM to NVM and vice-versa is equivalent to Equation~\ref{eq:dram} and Equation~\ref{eq:sttram}, respectively. The major difference is instead of counting the total accessed volume, here we only consider the size of the object and the migration cost in terms of time deemed with DRAM REF energy. Due to space limitations, we excluded the equation representation.

Using Equations~\ref{eq_energy_dton_total} and \ref{eq_energy_ntod_total}, the total amount of energy consumption involving object migration can be presented in Equation~\ref{eq_energy_left}. 
\begingroup\makeatletter\def\f@size{7.5}\check@mathfonts
\begin{equation} \label{eq_energy_left}
	\begin{split}
	E_{total} = \sum_{i=1}^{N}\  \biggl[ \  X_i\cdot \{CP_i\cdot dnE_i + (1-CP_i)\cdot ndE_i\} \\
	+ (1-X_i)\cdot \{CP_i\cdot dE_i + (1-CP_i)\cdot nE_i\} \  \biggl]
	\end{split}
\end{equation}

Equation~\ref{eq_energy_left} is the left-hand side of the energy limit constraint inequality. 
In the meantime, the right-hand side of the inequality may vary according to the purpose of the external request. The requested energy constraint can be categorized into two possibilities. First, the new energy constraint is effective only when the limit is strictly kept. That is, if the object migration cannot satisfy the new energy constraint, \emdyn{} does not shuffle the current placement of memory objects. Second, the new energy constraint does not require tight limiting. For instance, user may require to reduce energy consumption regardless of meeting the energy constraint. In this case, \emdyn{} shuffles the memory objects. 

To consider these different demands, the Migration Planner provides an additional flag, $F$, as an input parameter whose value is 1 when the purpose belongs to case (i) and 0 otherwise. By considering these cases, the requirement $Rq$ can be shown as Equation~\ref{eq_energy_right}.
\begingroup\makeatletter\def\f@size{7.5}\check@mathfonts
\begin{equation} \label{eq_energy_right}
	\begin{split}
	Rq = F\cdot \bigl[ \ \sum_{i=1}^{N}dE_i\cdot R_{n} \ \bigl]\ 
	+ \  (1-F)\cdot \bigl[ \ \sum_{i=1}^{N} \{CP_i\cdot dE_i \\+ (1-CP_i) \cdot nE_i \}\ \bigl]
	\end{split}
\end{equation}

Equation~\ref{eq_energy_right} becomes the right-hand side of the energy limiting inequality and $R_{n}$ indicates the newly required energy constraint. 
Therefore, the total energy constraint shown in Equation~\ref{eq_energy_left} should be less than and equation to the required energy shown in Equation~\ref{eq_energy_right}.

\paragraph{\textbf{Objective Function}}
The Migration Planner aims to minimize the memory access latency, and it can be calculated by the sum of total latency due to objects that are either migrated or not. 
If an object which is assigned to DRAM currently is a migration candidate to NVM, the total latency $(dnL_i)$ of that object can be shown as Equation~\ref{eq_objective_dton_total}. 
\begingroup\makeatletter\def\f@size{7.5}\check@mathfonts
\begin{equation} \label{eq_objective_dton_total}
	\begin{split}
	dnL_i = L_{D}\cdot L3M_i\cdot \frac{t-sT_i}{T_i} + MigCT_i + L_{N}\cdot L3M_i\cdot \frac{fT_i-t}{T_i}
	\end{split}
\end{equation}

The Placement Planner of \emplan{} profiles the number of LLC misses ($L3M_I$) of memory object in advance to count the number of actual memory accesses. However, the information of how many LLC misses would occur during the object migration cannot be measured before runtime. Thus, we assume that the access to the memory device would occur for the whole object both in reading and writing. The migration cost ($MigCT_i$) for the time taken to migrate can be calculated by considering the access latency of each device ($L_{D}$ \& $L_N$) and the size of the memory object.

Likewise, an object which was placed to NVM previous is the migration candidate then its total access latency can be presented as Equation~\ref{eq_objective_ntod_total}\cb{.}
\begingroup\makeatletter\def\f@size{7.5}\check@mathfonts
\begin{equation} \label{eq_objective_ntod_total}
	\begin{split}
		ndL_i = L_{N}\cdot L3M_i\cdot \frac{t-sT_i}{T_i} + MigCT_i + L_{D}\cdot L3M_i\cdot \frac{fT_i-t}{T_i}
	\end{split}
\end{equation}	

Thus, the total delay time of the objects for migration can be presented as shown in Equation~\ref{eq_objective_total}.
\begingroup\makeatletter\def\f@size{7.5}\check@mathfonts
\begin{equation} \label{eq_objective_total}
	 \begin{split}
		f = \sum_{i=1}^{N}\  \biggl[ \  X_i\cdot \{CP_i\cdot dnL_i + (1-CP_i)\cdot ndL_i\} + \\(1-X_i)\cdot L3M_i\cdot \{L_{D}\cdot CP_i + L_{N}\cdot (1-CP_i)\} \  \biggl]
  	\end{split}
\end{equation} 

\subsubsection{Migration Executor} \label{migrate-exector}
Once the migration decision for all the target objects is made, the Migration executor operates to perform the migration task and relocate the memory objects between respective memory modules.
The steps of the object migration are as follow:
\begin{itemize}
	\item A new object with the same size of the candidate object is allocated in the respective memory module.
	\item The currently stored data of the candidate object is copied to the new object.
	\item The pointer of the candidate object is revised to point towards the newly allocated object.
	\item The candidate object is de-allocated.
\end{itemize}

In step 3, Migration executor should maintain the address values of not only the pointer directly referring to the object in the allocation time, but also all general pointer variables which point to the object in the target application. 
In this work, we implement a member function that registers application pointers' addresses, and we have called it in every pointer reference on major objects in the target application. But, this method incurs application code modification to register the pointer addresses for Migration executor. To deal with this problem, the proxy pointer concept, which is similar to the proxy object suggested by~\cite{nvheap}, can be applied. By maintaining one proxy pointer per major object, Migration executor can set all application pointers to refer to this proxy pointer. Migration executor will be able to migrate the objects only with changing the destination of proxy pointer. In case of migration, two minor issues that need to be consider during the migration are the migration scenarios and the case of failure at the migration. 
{\paragraph{Migration Cases:}} 
Some of the example of migration cases can be: (1) when the user deliberately wants energy efficiency due to high charges of supporting systems from the data-center service providers and (2) when the system is required to reduce the energy consumption for long running applications to provide resources to the other applications. 
{\paragraph{Failure-safe Migration:}} 
The memory objects are migrated in a failure-safe manner across the memory modules. For instance, if the failure occurs at Step 2 of the migration executor, the application will still access the previous object pointer as the pointer in the application is not updated or if the failure occurs during the Step 3 of the migration executor, the application will still access the previous pointer as the new pointer is not updated completely. 

\begin{table}[!b]
	\caption[Test-bed specifications and benchmark workloads]{Testbed specifications and benchmark workloads } 
	\label{tab:table_environ}
	\resizebox{\columnwidth}{!}{
	\begin{tabular}{l|l|ll}
			\hline
			\textbf{Configuration} & \textbf{Component} & \textbf{Value} \\ \hline
			\multirow{7}{*}{Test-bed I} & Processor & Intel Xeon 
E5-2650V4, 2 Sockets, 8 cores \\&&per socket &\\ 
			& L1 Cache & 32KB 8-way set-associative (per core) &\\ 
			& LLC & 20MB 8-way set-associative (shared) &\\ \
			& Memory & 2 channel, 16GB, 16 banks, 16KB row buffer &\\
			& DRAM Latency & Read: 200 (ns), Write: 200 (ns) &\\
			& STT-RAM Latency & Read: 640 (ns) Write: 1440 (ns) &\\ \hline
			\multirow{7}{*}{Test-bed II} & Processor & Intel Xeon CPU E5-4640, 4 Sockets, 10 cores \\&&per socket & \\ 
			& L1 Cache & 64KB 8-way set-associative (per core) &\\
			& LLC & 20MB 8-way set-associative (shared) &\\ \
			& Memory & 2 channel, 8GB, 16 banks, 16KB row buffer &\\ 
			& DRAM Latency & Read: 400 (ns), Write: 400 (ns) &\\
			& STT-RAM Latency & Read: 840 (ns) Write: 1640 (ns) &\\ \hline
			NVM Emulation & Emulation Tool & Quartz &\\ \hline
			\multirow{5}{*}{Benchmark}& Benchmark & NPB~\cite{Bailey2011}, PBBS~\cite{pbbs} &\\
			& Applications & NPB: Conjugate Gradient (CG), Fourier Transform (FT) \\&&PBBS: Breadth First Search (BFS), Spanning Forest (SF) &\\ 
			& Memory Footprint & NPB: CG: 1.08GB, FT: 1.08GB \\&&PBBS: BFS: 6.78GB, SF: 4.3GB &\\\hline
    \end{tabular}} 
\end{table}

\section{Evaluation}\label{sec:eval}

\subsection{Experimental Setup}
We evaluate our proposed \emap{} system on two different testbed configurations and we have evaluated two benchmarks, Problem-based Benchmark Suit (PBBS)~\cite{pbbs} and NAS Parallel Benchmark (NPB)~\cite{Bailey2011} as shown in Table~\ref{tab:table_environ}. 

For the emulation of NVM in HMMS, we adopted the 
QUARTZ emulation platform~\cite{quartz}. The read and write latency of DRAM is considered as 10ns~\cite{7208275} while the read latency of STT-RAM is 32ns and write latency is 72ns~\cite{sttram_latency}. In our system configurations, the memory latency before emulation is measured to be 200 ns with QUARTZ~\cite{quartz}. We computed the ratio of DRAM to STT-RAM latency and shown in Table~\ref{tab:table_environ} for emulation. For the evaluation of the energy consumption in both testbed configurations, we calculated the estimated energy consumption with suggested equations. Equation variables (memory access patterns) are derived from object-level profiling. For the evaluation, we only present the estimated energy consumption of the memory system excluding the CPU and caches. As measuring the energy consumption of memory systems in real-time is not possible due to lack of measuring tools. 

We compared \emplan{} with MOCA~\cite{moca}, which improves both performance and energy by selectively placing memory objects in HMMS.
MOCA measures the LLC MPKI of objects in HMMS consisting of high-bandwidth, low-latency, and low-power memory modules. In addition, MOCA also considers the memory-level parallelism in profiling which is beyond the scope of this work. It allocates memory-intensive objects which have high LLC MPKI values to high-bandwidth and low-latency memory modules. This methodology is applicable to HMMS that composed of DRAM and NVM by considering DRAM as high-bandwidth and low-latency memory. 

\subsection{\lowercase{e}Mplan Performance and Energy Evaluation}
In this section, we will present the performance and energy estimation evaluation of our static module of \emap{}, \emplan{}, using the PBBS and NPB on Testbed I. \subsubsection{Analysis on PBBS Applications (BFS, SF)}
In this section, we compare the results of \emplan{} placement with multiple energy limiting constraints by counter-part placement methodology, MOCA~\cite{moca}. 
\begin{figure}[t]
	\centering
	\subfloat[Execution time of BFS]{
	  \includegraphics[width=0.4\textwidth]{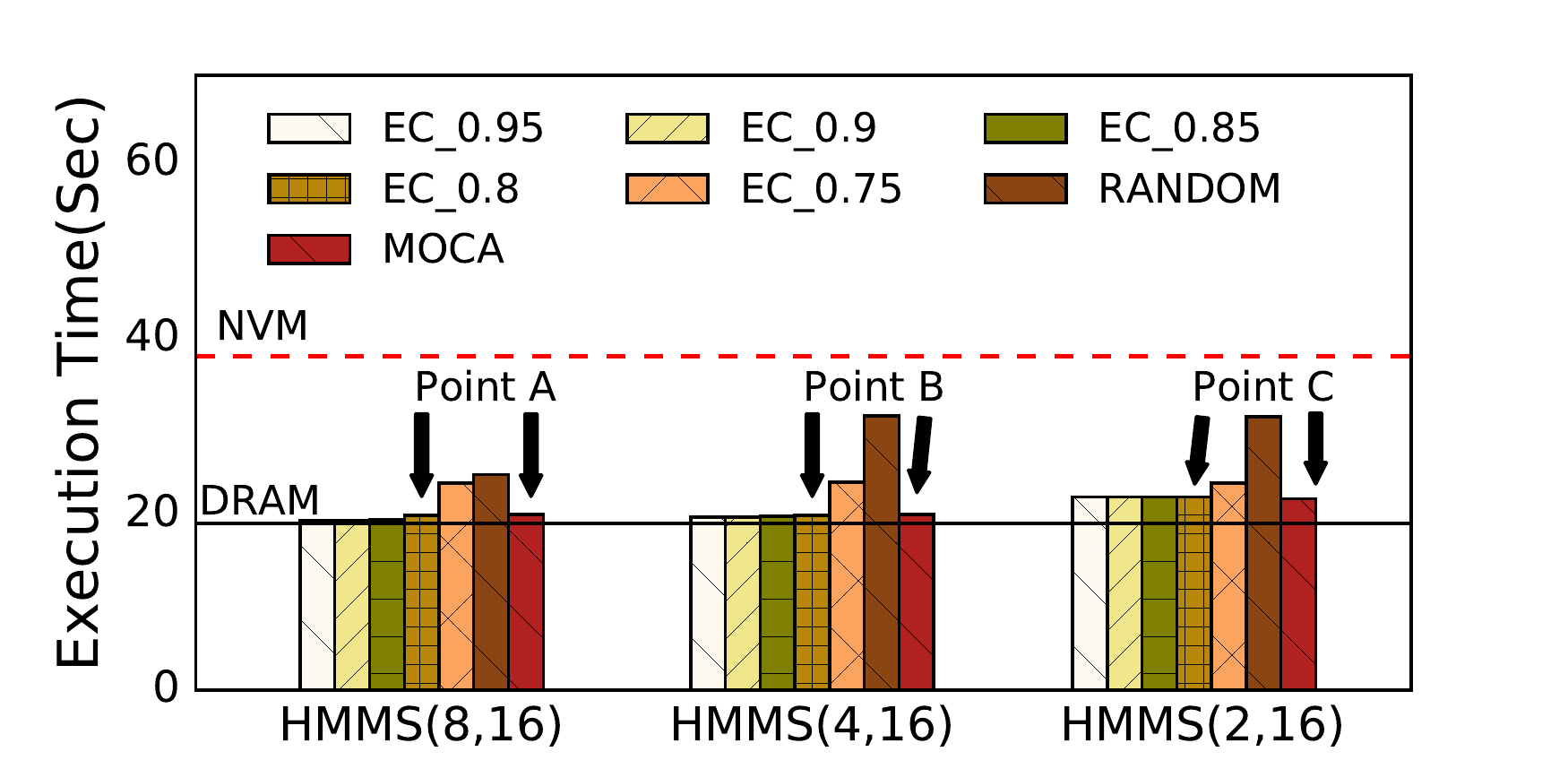}%
	}

	\subfloat[Estimated energy of BFS]{%
	  \includegraphics[width=0.4\textwidth]{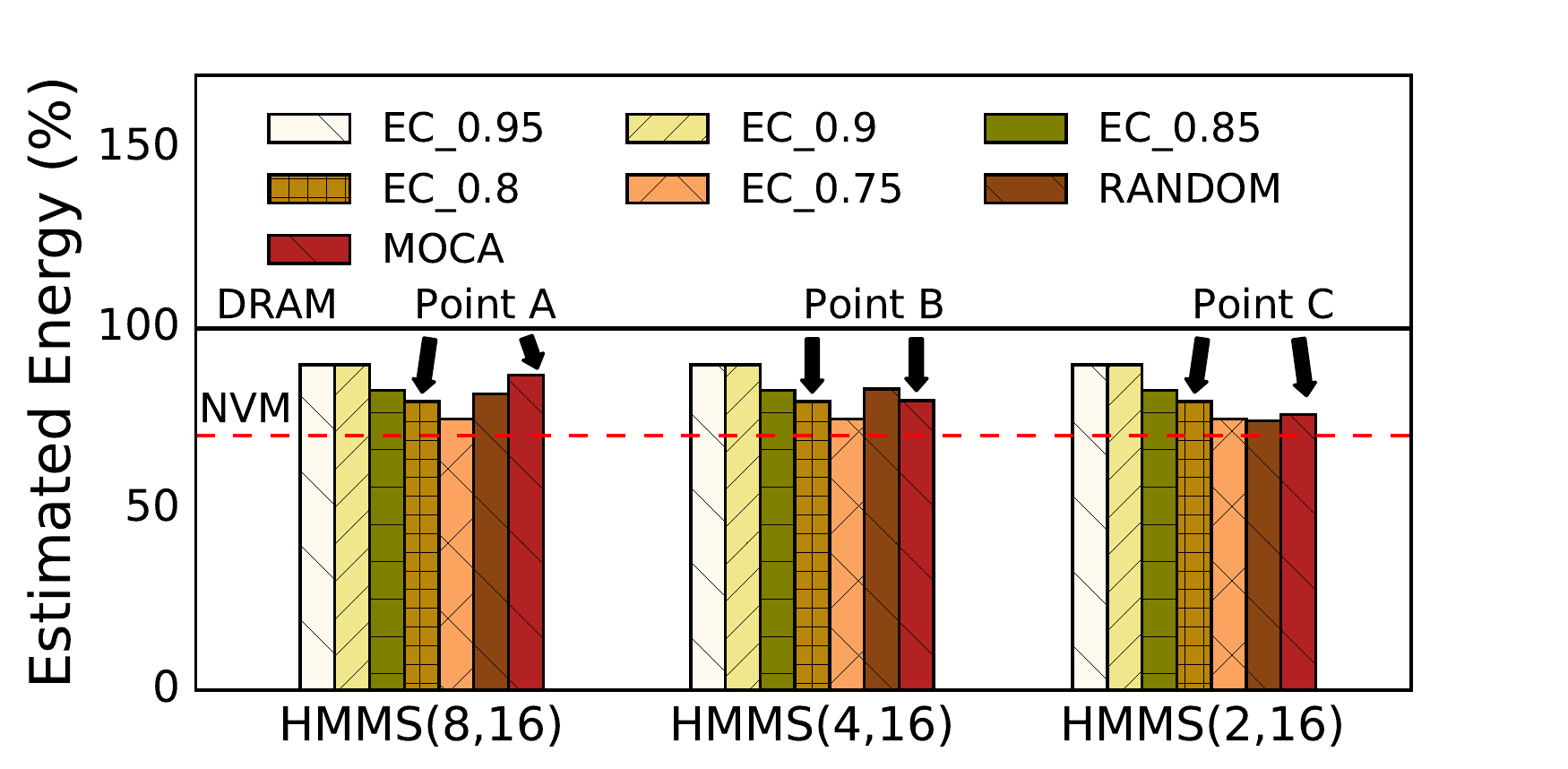}%
	}
	\caption{ \small The performance and energy consumption of the PBBS BFS Application. The x-axis represents different HMMS configurations in terms of capacities of DRAM and NVM. HMMS(8, 16) determines that DRAM is 8 GB while NVM is 16 GB. While y-axis shows the execution time and estimated energy consumption percentage for both, respectively.  }
	\label{fig:pbbs_bfs}
\end{figure}

\begin{figure}[t]
	\centering
	\subfloat[Execution time of SF]{
	  \includegraphics[width=0.4\textwidth]{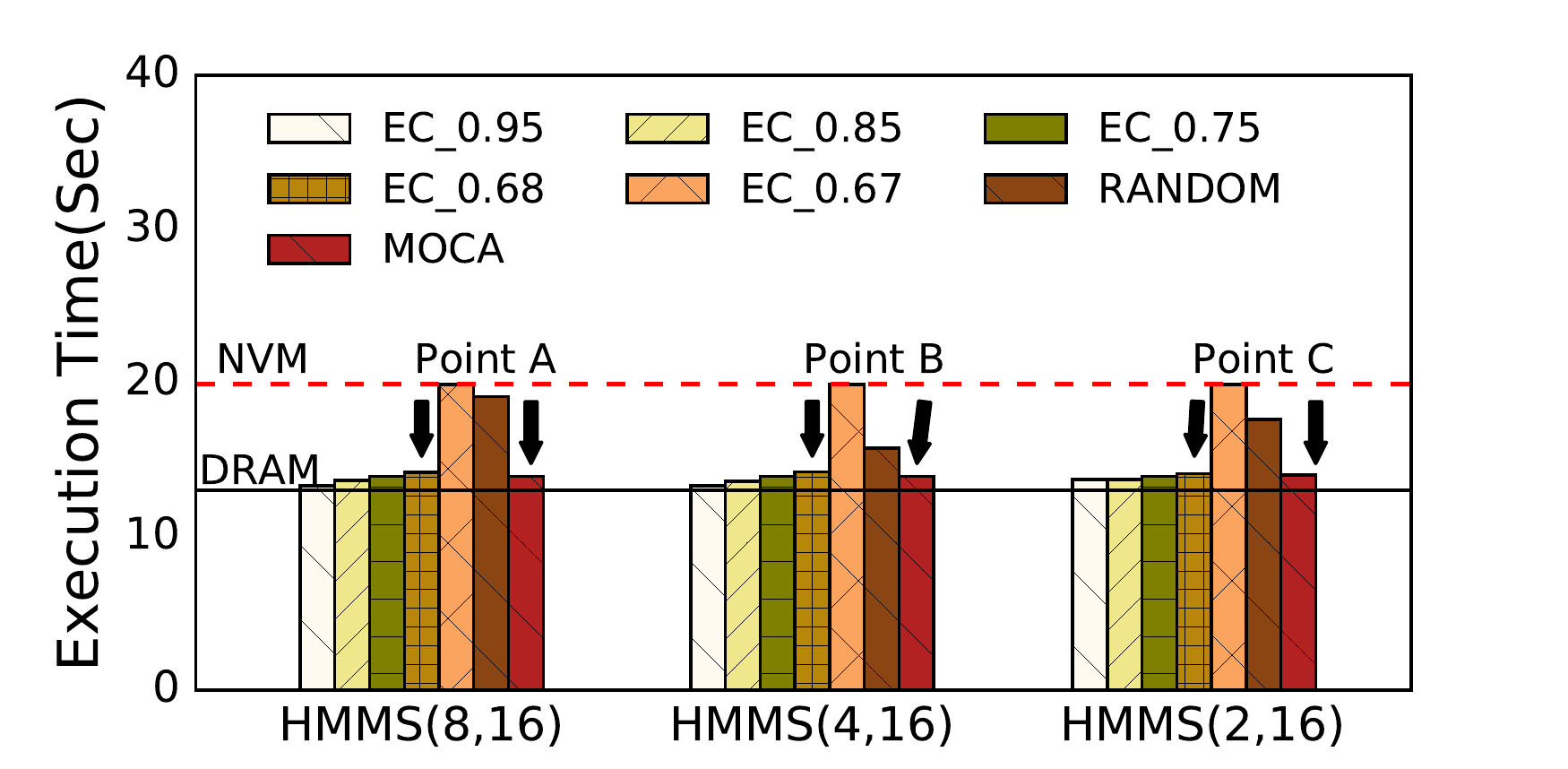}
	}

	\subfloat[Estimated energy of SF]{
	  \includegraphics[width=0.4\textwidth]{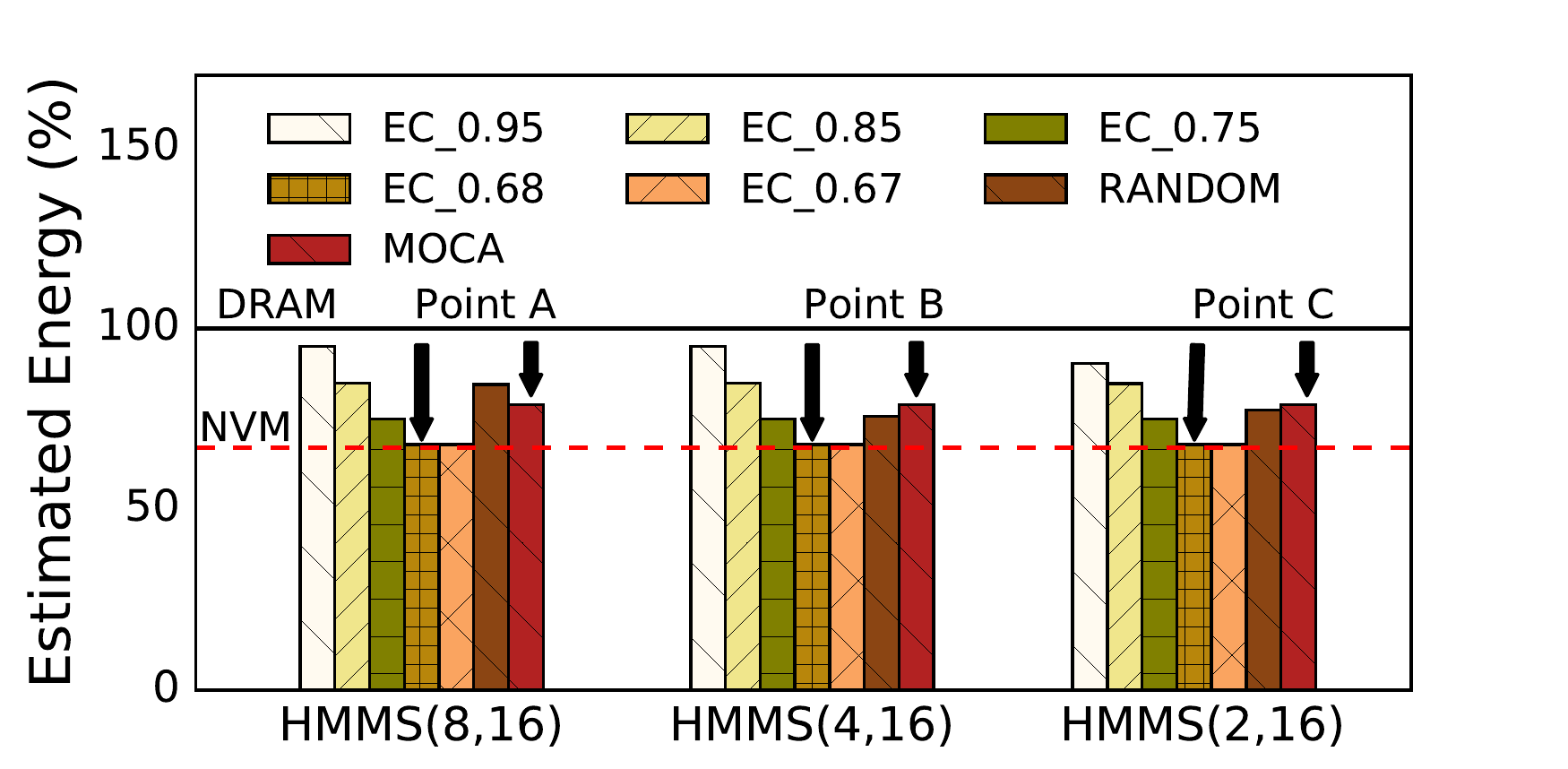}
	}
	\caption{\small The performance and energy consumption of the PBBS SF Application. The x-axis represents different HMMS configurations while y-axis shows the execution time and estimated energy consumption percentage for both, respectively.}
	\label{fig:pbbs_sf}
\end{figure}

Figure~\ref{fig:pbbs_bfs}(a) and (b) show the performance and estimated energy consumption of the proposed \emplan{} for BFS application, respectively. Considering the larger density of NVM, we conduct various experiments while reducing the capacity of DRAM in HMMS. The memory footprint of the workload is shown in Table~\ref{tab:table_environ} and the selection of HMMS configuration was from extreme limited to enough capacity in terms of DRAM. In Figure~\ref{fig:pbbs_bfs}, the $EC\_X$ shows the energy limiting constraint in contrast to DRAM-only that the allocated objects will not exceed $X$ times of the energy consumption. For example, $EC\_0.9$ is the case where objects are allocated in HMMS to consume energy less than 90\% of the DRAM-only case and so on. 
On the other hand, the random case shows the execution time and estimated energy consumption when objects are randomly allocated without any placement decision in a range which does not exceed the capacity of given memory devices. 

The MOCA in the experimental results are the object allocation followed by the methodology~\cite{moca}. In Figure~\ref{fig:pbbs_bfs}(a), the execution time increased as the energy constraint becomes more restricted such as $EC\_0.8$ and above. The reason is \emplan{} gives priority to performance-critical objects to be placed on DRAM and once the energy limit constraint becomes more strict than 80\%, it starts to allocate the performance-critical objects to NVM. Nevertheless, if an application user wants to sacrifice some performance to reduce more energy, one may need intense constraint over 80\%. Random and MOCA methodologies cannot consider the placement which decreases further energy consumption with performance trade-off. Figure~\ref{fig:pbbs_bfs}(b) shows that \emplan{} meets the given energy constraints. The random placement has shown the worst energy efficiency where its execution time is longer than $EC\_0.75$. 

Figure~\ref{fig:pbbs_bfs}(b) has also shown that \emplan{} is more energy-efficient than the MOCA methodology. The A, B, and C points in Figure~\ref{fig:pbbs_bfs} show that the placement policies of \emplan{} and MOCA are almost similar, however, at point A, the energy consumption of MOCA is 8.2\% higher than \emplan{}. In contrast, at point B the performance and energy consumption of both are almost identical. In addition, at point C, the energy consumption of MOCA is less than \emplan{} as the proposed approach prefers to place performance critical objects on DRAM to optimize the performance of HMMS. On the other hand, MOCA only provides one-time placement policies for memory objects without considering the user requirements for performance and energy efficiency. 
\begin{figure}[t]
	\centering
	\subfloat[\emplan{} $EC\_0.8$]{\includegraphics[width=0.4\textwidth]{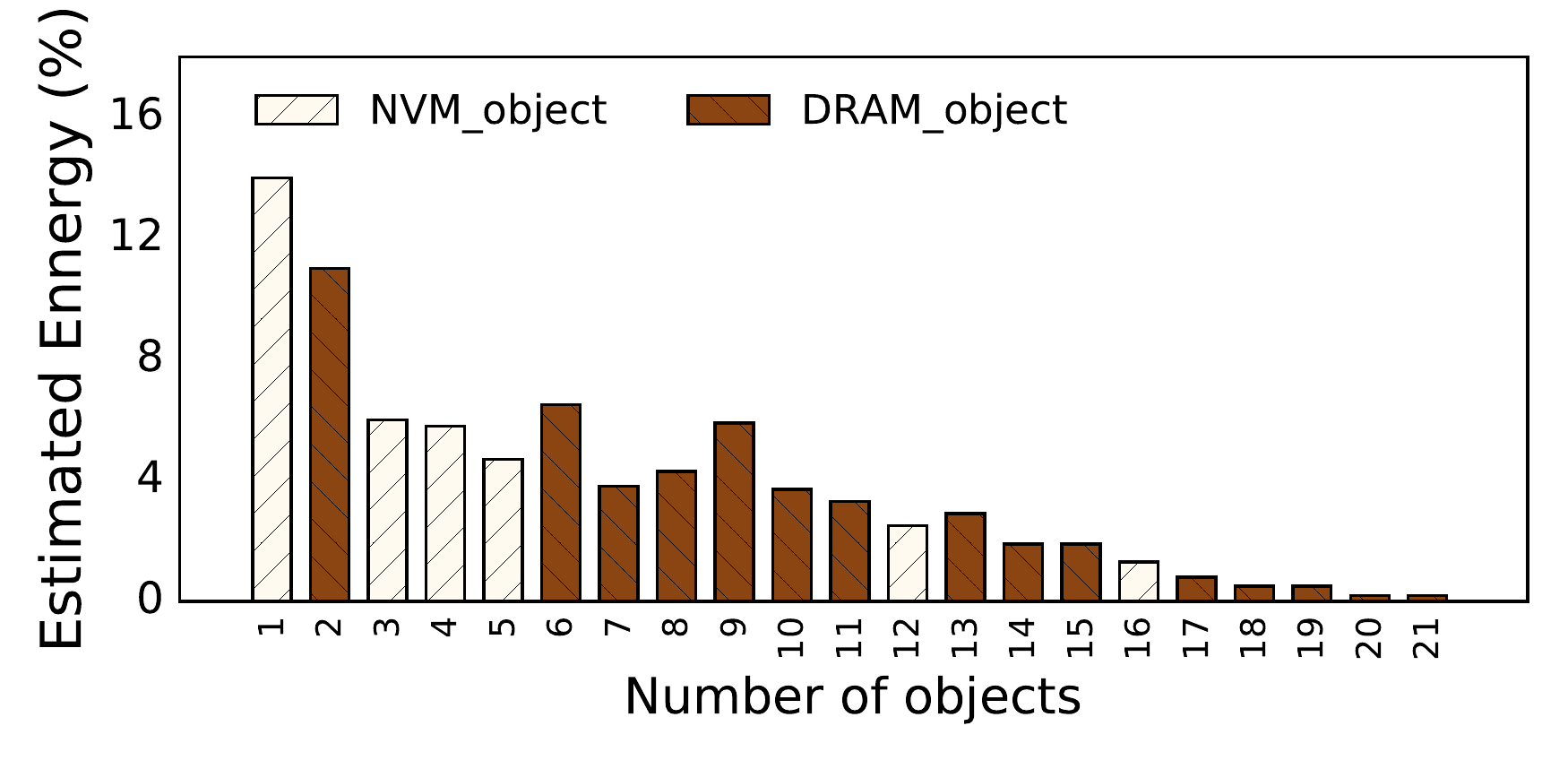}}
	
	\subfloat[MOCA]{\includegraphics[width=0.4\textwidth]{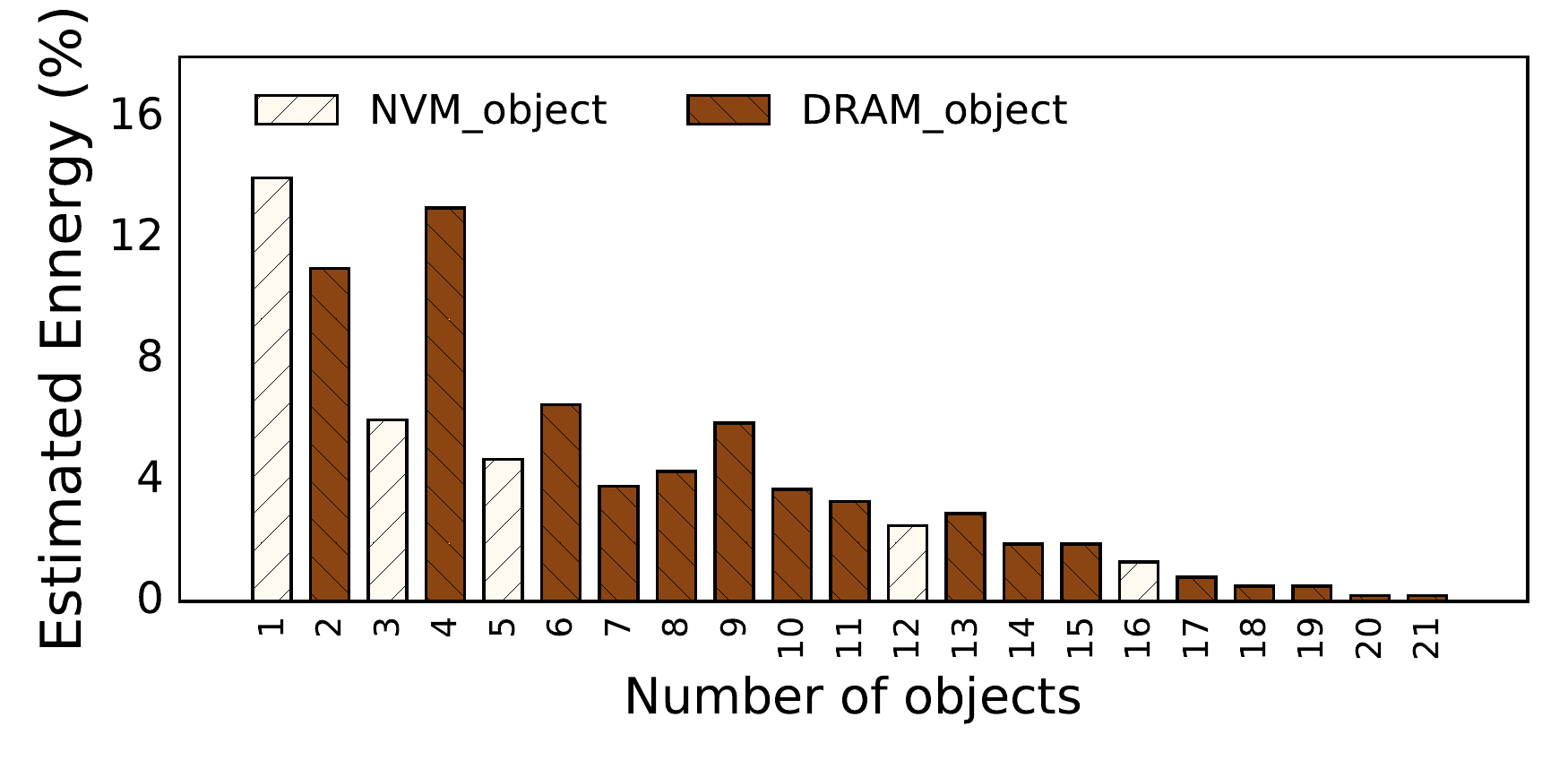}}
	\caption{ \small Per-object energy consumption of PBBS BFS (normalized to all-DRAM energy consumption). The x-axis shows the number of memory objects while y-axis is the estimated energy consumption.} 
	\label{fig_eval_bfs_obj_E}
\end{figure}

To better understand this, we analyze the per-object energy consumption of the BFS application as shown in Figure~\ref{fig_eval_bfs_obj_E}. The object placement decision of both techniques is almost similar except $4$-th object, where \emplan{} has placed that object in NVM while MOCA has placed it on DRAM. $4$-th object has the second-longest lifetime among objects of BFS and occupies the largest memory usage (826 MB), so if it is placed to DRAM, the amount of energy consumed in refreshing is large. Also, though the LLC MPKI value of $4$-th object is bigger than the threshold (0.025), it is not too large enough to impact performance. Therefore, when $4$-th object is allocated to DRAM, it does not only result in significant performance improvement but also consumes over 2.2x more energy. The \emplan{} module places $4$-th object to NVM by considering this in advance, but MOCA places it in DRAM because MOCA cannot consider object access pattern and memory devices characteristics. 

Spanning Forest (SF) of PBBS benchmark also shows consistent results with BFS. 
Figure~\ref{fig:pbbs_sf}(a) and (b) show the performance and estimated energy consumption at given energy constraint. Figure~\ref{fig:pbbs_sf}(a) shows that the execution of \emplan{} is same until the EC\_0.67 as the application latency increases as the energy constraint go beyond 67\% of DRAM-only. This is because \emplan{} effectively works to minimize the latency while satisfying the energy constraint until 68\% of DRAM-only as it place the latency-insensitive objects to NVM in order to further reduce the energy consumption on 67\% of DRAM-only energy constraint. Figure~\ref{fig:pbbs_sf}(a) and (b) also show that \emplan{} is more energy efficient than MOCA. The A, B, and C points in Figure~\ref{fig:pbbs_sf} show that \emplan{} placement decisions at energy constraint EC\_0.68 have the same application execution time as MOCA. However, the estimated energy consumption is 14\% more efficient than MOCA as \emplan{} places the memory objects by considering detailed access patterns and the characteristics of memory devices. Thus our methodology places only those objects to NVM which has better energy efficiency than MOCA. On the other hand, MOCA only considers the Last-Level Cache misses and memory-level parallelism to decide the placement of memory objects which leads to sub-optimal memory placement decisions and results in high energy consumption consequently. 
\begin{figure}[t]
	\centering
	\subfloat[Execution time of CG]{
	  \includegraphics[width=0.4\textwidth]{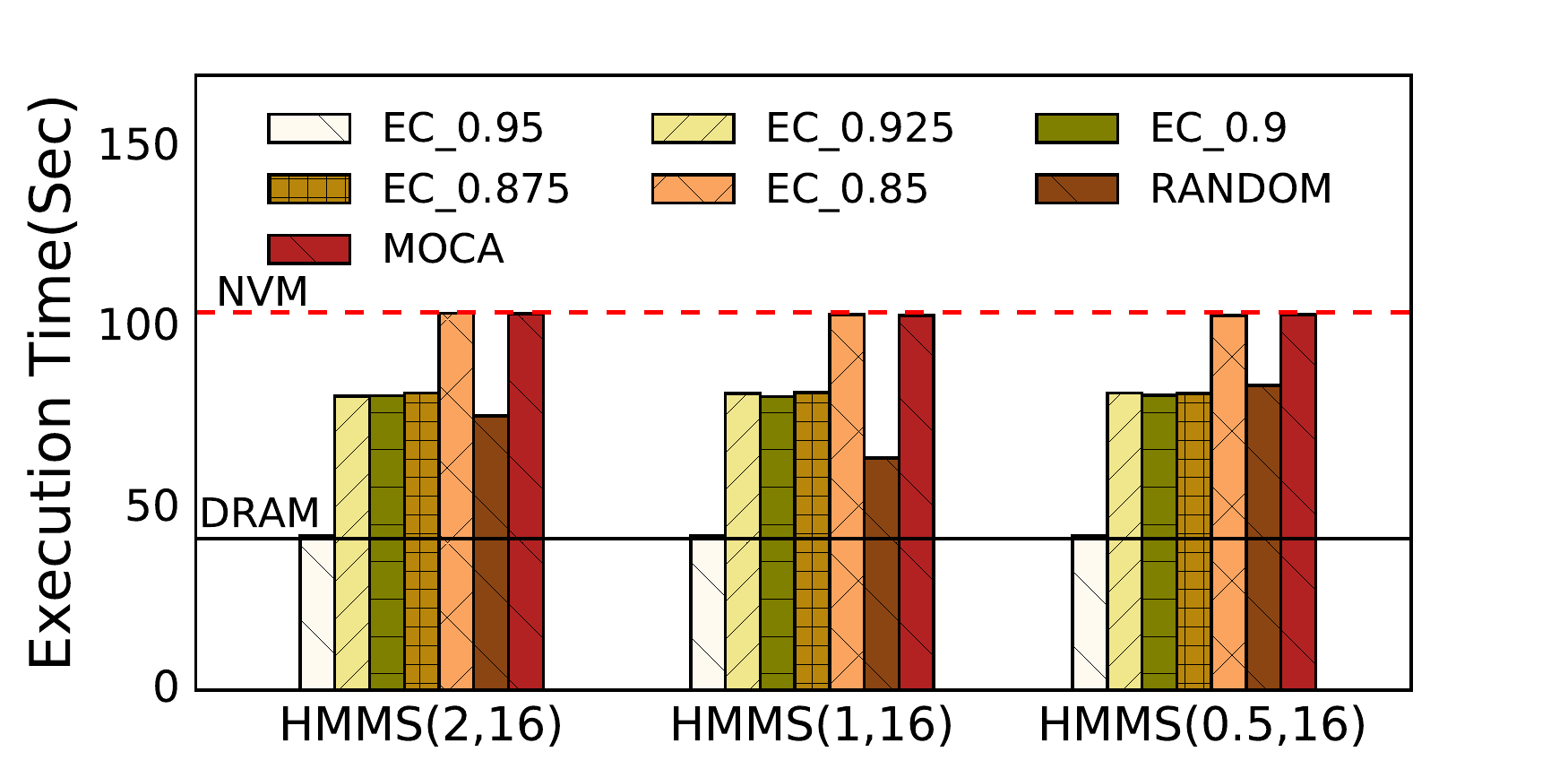}
	}

	\subfloat[Estimated energy of CG]{
	  \includegraphics[width=0.4\textwidth]{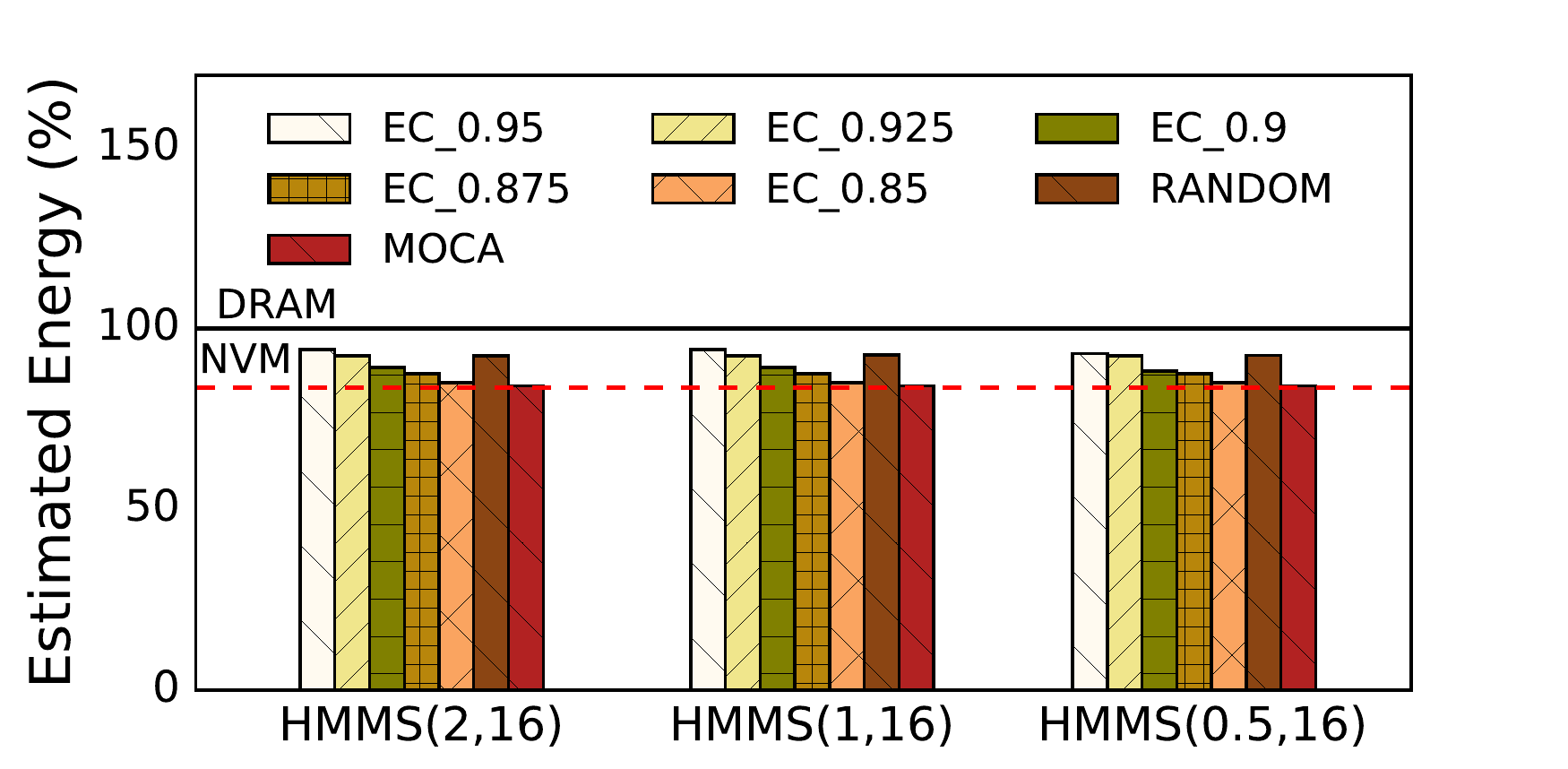}
	}
	\caption{\small The performance and energy consumption of the NPB CG Application. The x-axis represents different HMMS configurations while y-axis shows the execution time and estimated energy consumption percentage for both, respectively.}
	\label{fig:npb_cg}
\end{figure}

\subsubsection{Analysis of NPB Benchmark (CG, FT)}
We also evaluate the NPB benchmark, a high-performance computing workload, to analyze the results by changing energy limit constraints. The applications used in this experiment are CG and FT. Figure~\ref{fig:npb_cg}(a) and (b) show the performance and the estimated energy consumption of CG application with varying energy constraints. Fig~\ref{fig:npb_cg}(a) shows that performance deteriorates as the energy constraint is becoming strict due to the small number of objects that actually affect the performance. In CG, five out of 14 objects occupy almost 99\% of DRAM size. Thus, as energy constraint increases, major objects are placed in the NVM causing performance degradation. Figure~\ref{fig:npb_cg}(b) shows that all the placement methodologies satisfied the energy constraint. But for CG, the MOCA placement has the lowest energy consumption but the longest execution time. 
If a low energy limit and fast execution time are required, the current MOCA methodology cannot satisfy this requirement.

FT application of the NPB benchmark exhibits the same execution patterns as CG. 
Figure~\ref{fig:npb_ft}(a) shows the execution time of FT, as the energy constraint is becoming strict the application performance is degraded due to the limited number of major objects. FT has only six objects in total where four of them occupy 99.7\% of total DRAM space. After certain energy constraints, objects that have major impacts on the execution time must be placed to NVM in order to meet the required energy limit. Thus, when those objects are allocated to NVM, performance is decreased rapidly. Figure~\ref{fig:npb_ft}(b) shows that \emplan{} meets the given energy constraint. In FT, the placement policy of \emplan{} at energy constraint EC\_0.9 has a similar execution time as of MOCA in HMMS (2, 16) configuration while it has 4.3\% more energy efficiency. This is because when the DRAM capacity is sufficient, the \emplan{} module can take advantage of energy consumption by calculating the object placement which MOCA methodology cannot account for. However, when DRAM capacity is reduced, placement cases of \emplan{} are strictly limited, so the energy difference between \emplan{} and MOCA placements decreases. 
\begin{figure}[t]
	\centering
	\subfloat[Execution time of FT]{
	  \includegraphics[width=0.4\textwidth]{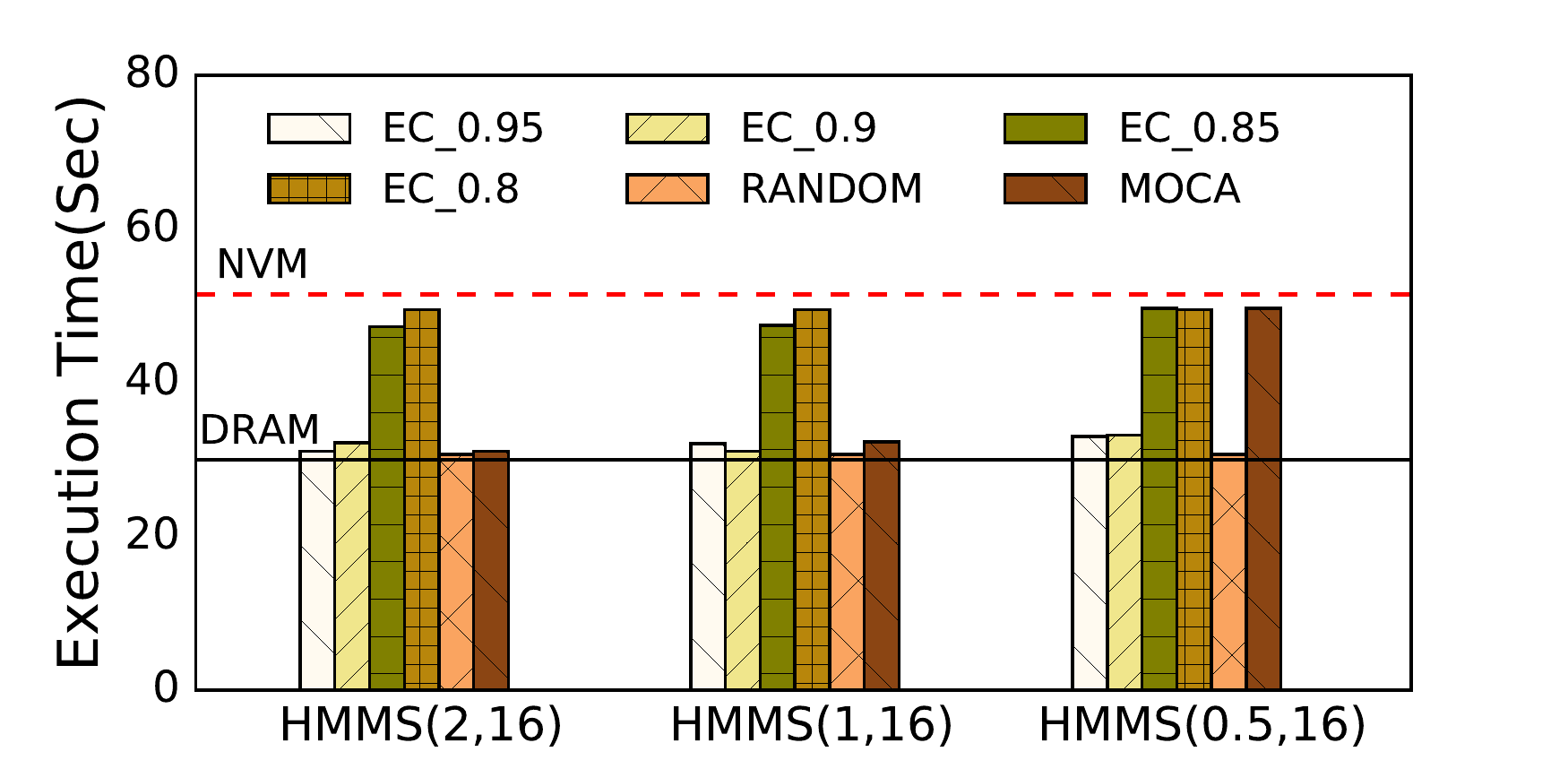}
	}

	\subfloat[Estimated energy of FT]{
	  \includegraphics[width=0.4\textwidth]{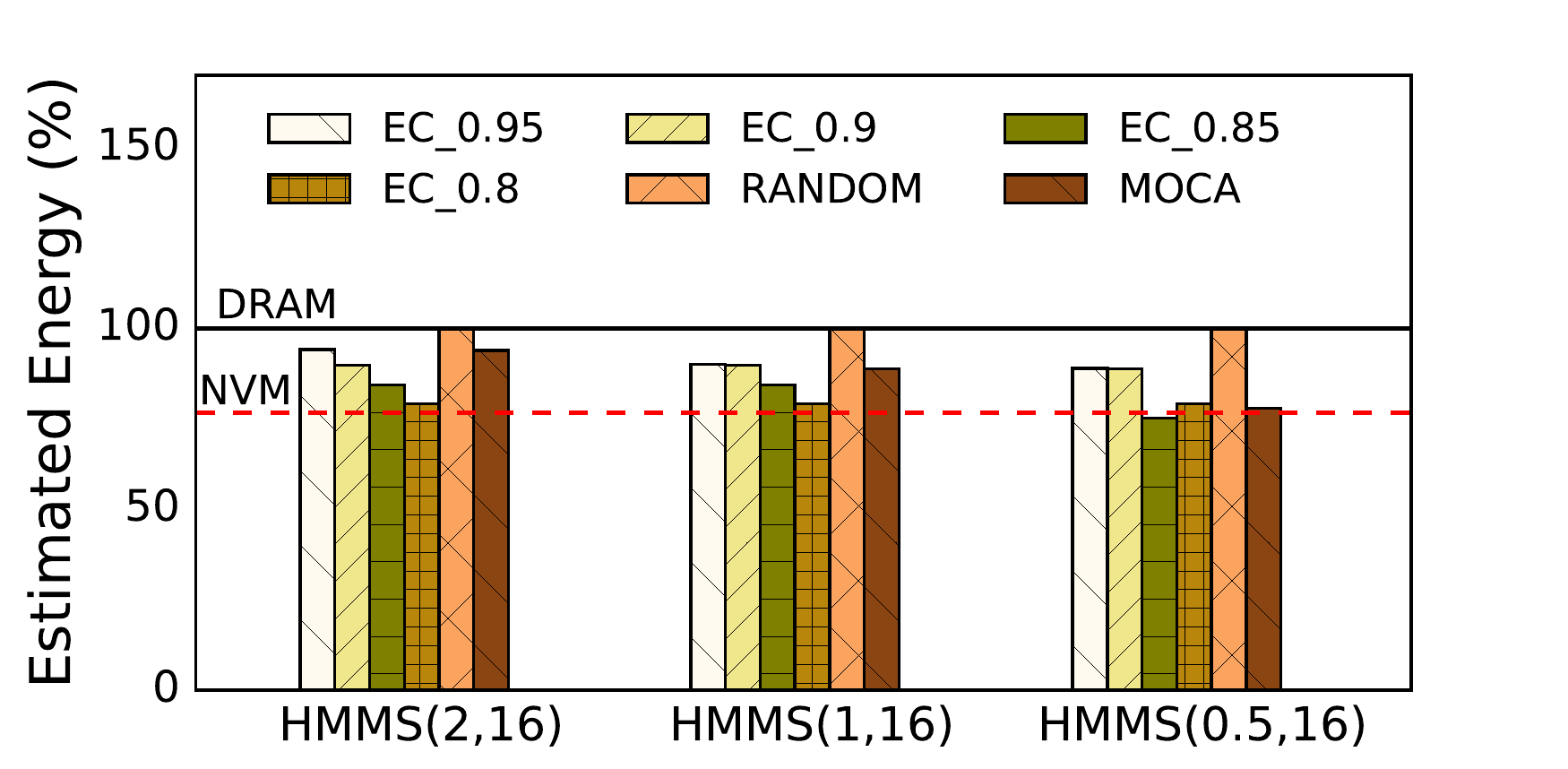}
	}
	\caption{\small The performance and energy consumption of the NPB FT application. The x-axis represents different HMMS configurations while y-axis shows the execution time and estimated energy consumption percentage for both, respectively.}
	\label{fig:npb_ft}
\end{figure}
\subsection{Energy Consumption Comparison MOCA vs \lowercase{e}mplan} 
\label{mova-eval}

In this experiment, we modified the MOCA methodology and configure it to meet the energy constraint. In original MOCA~\cite{moca}, objects are allocated on the basis of the specific threshold of the LLC MPKI, if the object has met the threshold then it will be placed in high performant memory otherwise placed at low-performing memory device. The LLC MPKI threshold is derived from several experiments for efficient performance while maintaining energy consumption. MOCA can also satisfy the energy limiting constraints if we set the LLC MPKI threshold effectively. However, MOCA cannot estimate the amount of energy consumed by each object in DRAM and NVM based binary HMMS because it does not consider the detailed object access patterns and NVM device characteristics. Through MOCA, the threshold to satisfy the energy constraint cannot be calculated but it must be empirically set by performing several experiments repeatedly.

MOCA samples the LLC MPKI and ROBH stall cycle information at a fixed interval (i.e. 1000 instructions) when the target application is executed and records the information along with the call-stack. After application execution is finished, MOCA maps those information to the object via allocation function call-stack, and then calculates the memory object energy consumption with per-object information. That is, if we assume that MOCA uses object access pattern profiling and energy models of DRAM and NVM in this research, MOCA can calculate the total energy consumption after the execution of the target application.
\begin{figure}[t]
	\centering
	\includegraphics[width=0.3\textwidth]{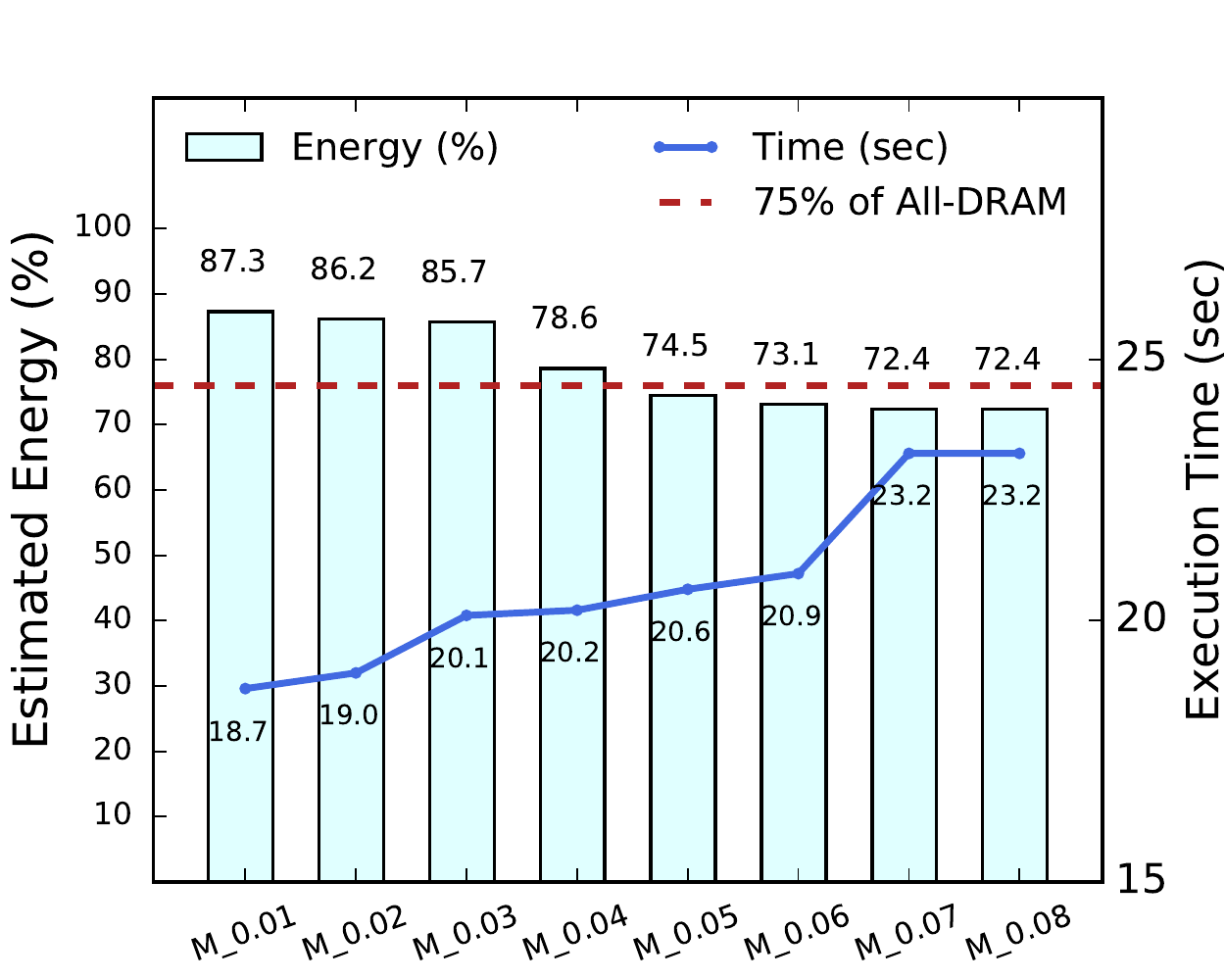}
	\caption{PBBS BFS Execution Time \& Energy Consumption Estimation of MOCA on Various LLC MPKI Values. (Estimated energy consumption is normalized to All-DRAM energy). The x-asis is the LLC MPKI threshold, M\_$Thr$. }
	\label{fig_eval_moca_energy}
\end{figure}

Figure~\ref{fig_eval_moca_energy} shows the estimated energy consumption based on various LLC MPKI values of MOCA placement in the PBBS BFS. It shows if the LLC MPKI threshold varies by same unit, then the change on energy consumption does not have any consistent pattern. Thus, to find the LLC MPKI value satisfying the energy limit constraint through MOCA, the $Thr$ should be searched by a certain unit. For example, to meet the energy constraint that consumes less than 75\% of energy to DRAM-only, MOCA should search by a certain unit increase in LLC MPKI. In our experimental environment, \emplan{} module takes up to 23.635 seconds of placement computation time and object allocator overhead, which is only related to BFS application. On the other hand, MOCA should execute the application four times, which takes 78 seconds, to find the adequate LLC MPKI threshold to meet the energy constraint. With including real execution time that takes 20.6 seconds in M\_0.05, MOCA placement spends 4.17x than \emap{} execution time in this example.

Also, there are cases where the fluctuation of LLC MPKI threshold value, which affects energy consumption and performance, is extremely minimal. In the case of NPB CG, for example, when the LLC MPKI threshold is 0.0024, the execution takes 68.715 seconds, and its energy consumption is equivalent to 91.9\% of DRAM-only. But, when the LLC MPKI threshold is lowered to 0.0023, the execution takes 42.473 seconds and consumes 95.1\% of energy of DRAM-only. If the energy consumption limit less than 92\% of DRAM-only is required, the effective LLC MPKI threshold can be obtained by performing LLC MPKI value search in units of 0.0001. 
When the search is performed in \cb{a} smaller unit, the search overhead increases accordingly and process need to be repeated every time the energy limit is changed.  
\begin{figure}[t!]
	\centering
	\subfloat[Execution time of the BFS]{
	  \includegraphics[width=0.4\textwidth]{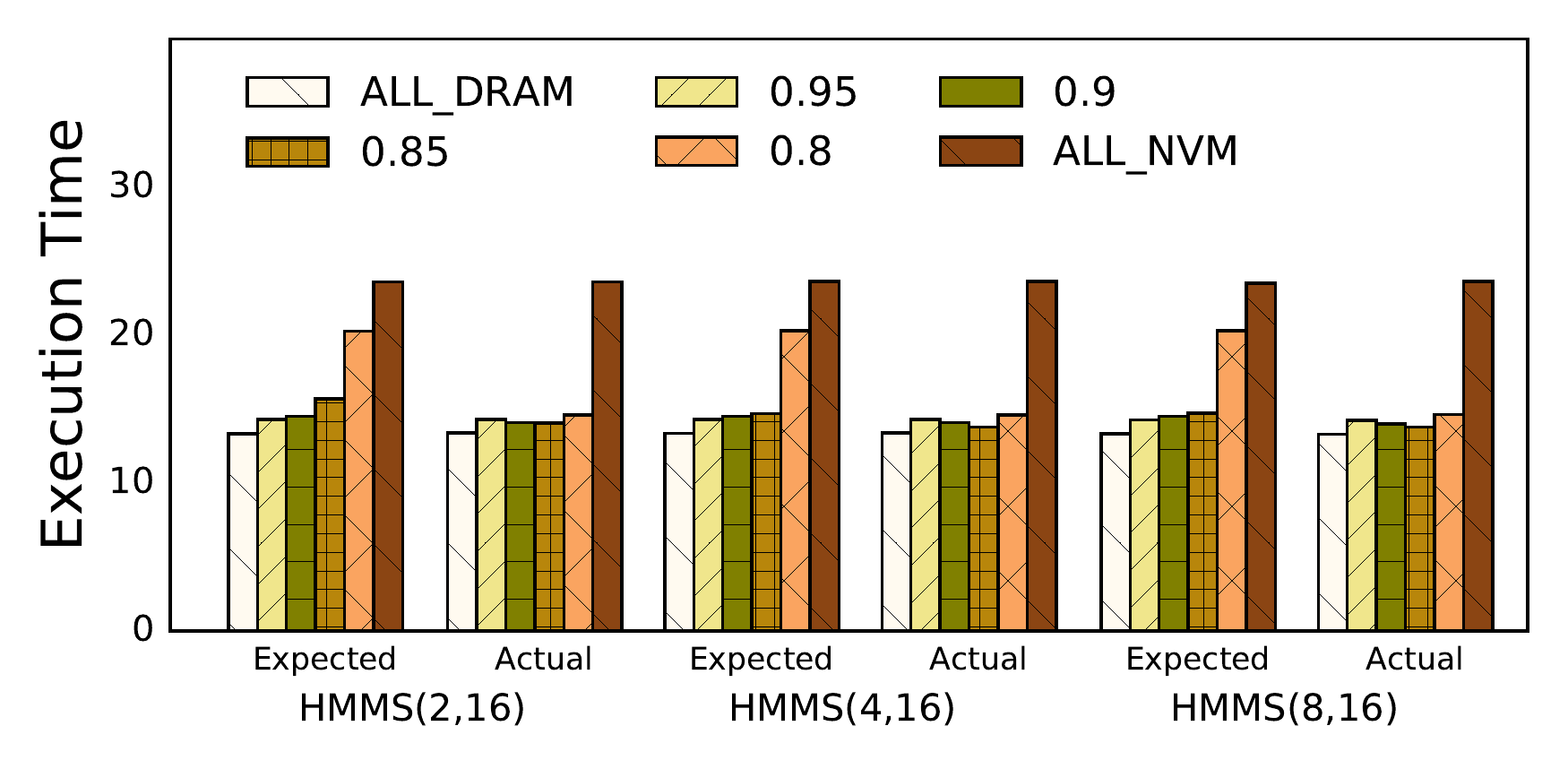}
	}

	\subfloat[Estimated energy of the BFS]{
	  \includegraphics[width=0.4\textwidth]{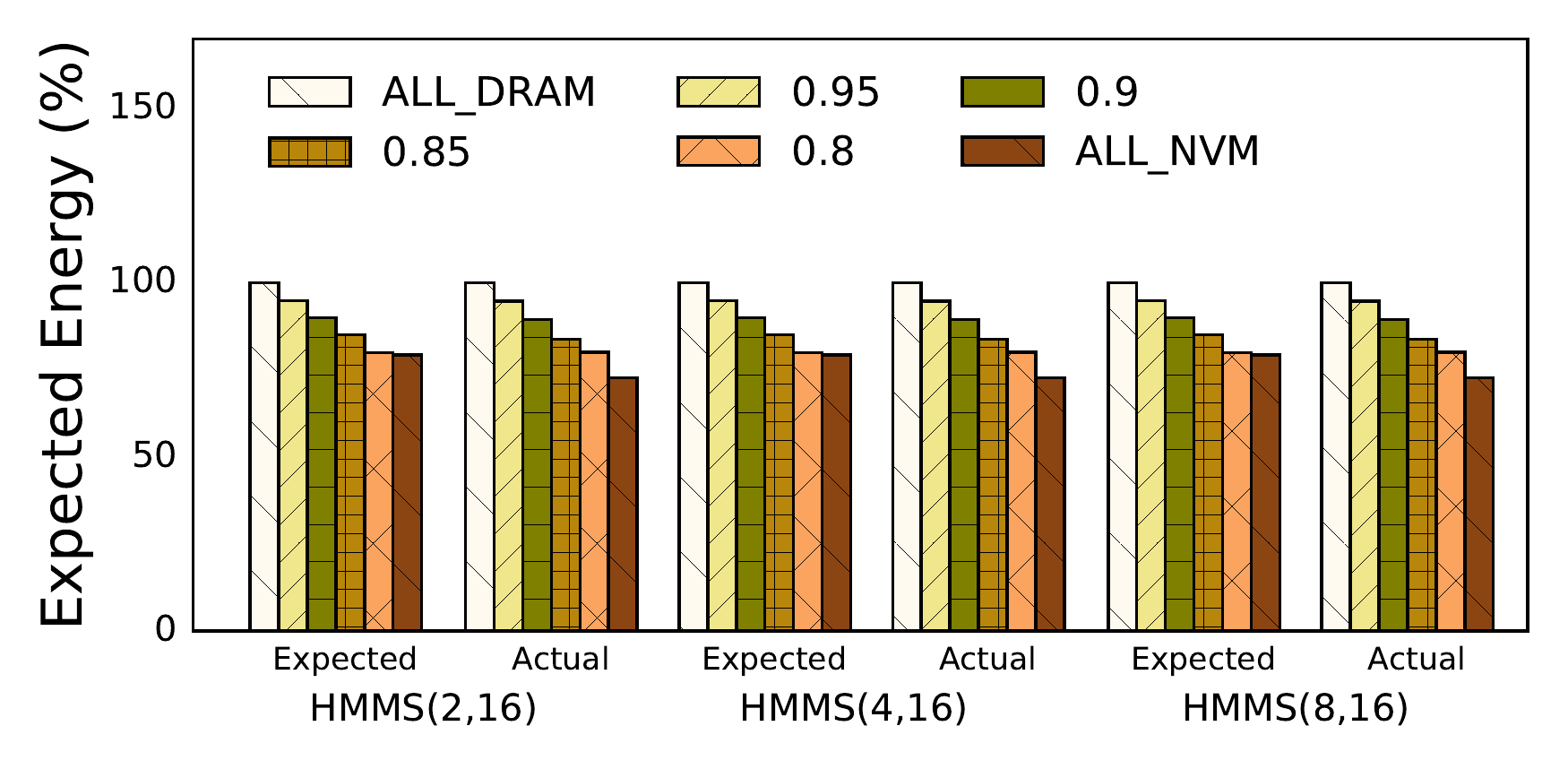}
	}

	\caption{\small The performance and energy consumption comparison with scaled and actual object placement policies. The x-axis shows the expected (computed using proposed scaling vector), actual (through profiling), and various HMMS configurations.}
	\label{fig:scale}
\end{figure}

\subsection{Accuracy of Scaling Rate Vector}
In this section, we evaluated the accuracy of our proposed scaling rate vector to avoid the profiling of the application whenever the workload changes. As the workload of an application varies, the access information also changes accordingly and application workloads can be categorized into three groups; fixed, scaling, and irregular~\cite{deepmap}. Most of the applications from scientific group lies in the scaling category as their access patterns scale with the scaling workload. For this experiment, we profiled the BFS application with various workloads and calculated the scaling rate vector for all the major variables as explained in section~\ref{scaling}. We present the accuracy of our proposed scaling rate vector in terms of the placement of memory objects in HMMS. We evaluated this experiment on Testbed II. 

BFS is categorized in the scaling class that as the input workload scales the access patterns of the variables also scales but the ratio of scaling is not consistent for most of the objects. So, we adopted the generalized way to calculate the scaling rate vector and shows the effectiveness of our proposal. 
Figure~\ref{fig:scale} show the performance and expected energy consumption with various HMMS configurations and energy constraints ($EC\_X$). The evaluation shows that most of the time it accurately places the memory object to their respective memory module, which ultimately omits the huge cost of profiling the application again with scaled workload. 
\begin{figure}[t]
	\centering
	\subfloat[Execution time of CG]{
	  \includegraphics[width=0.4\textwidth]{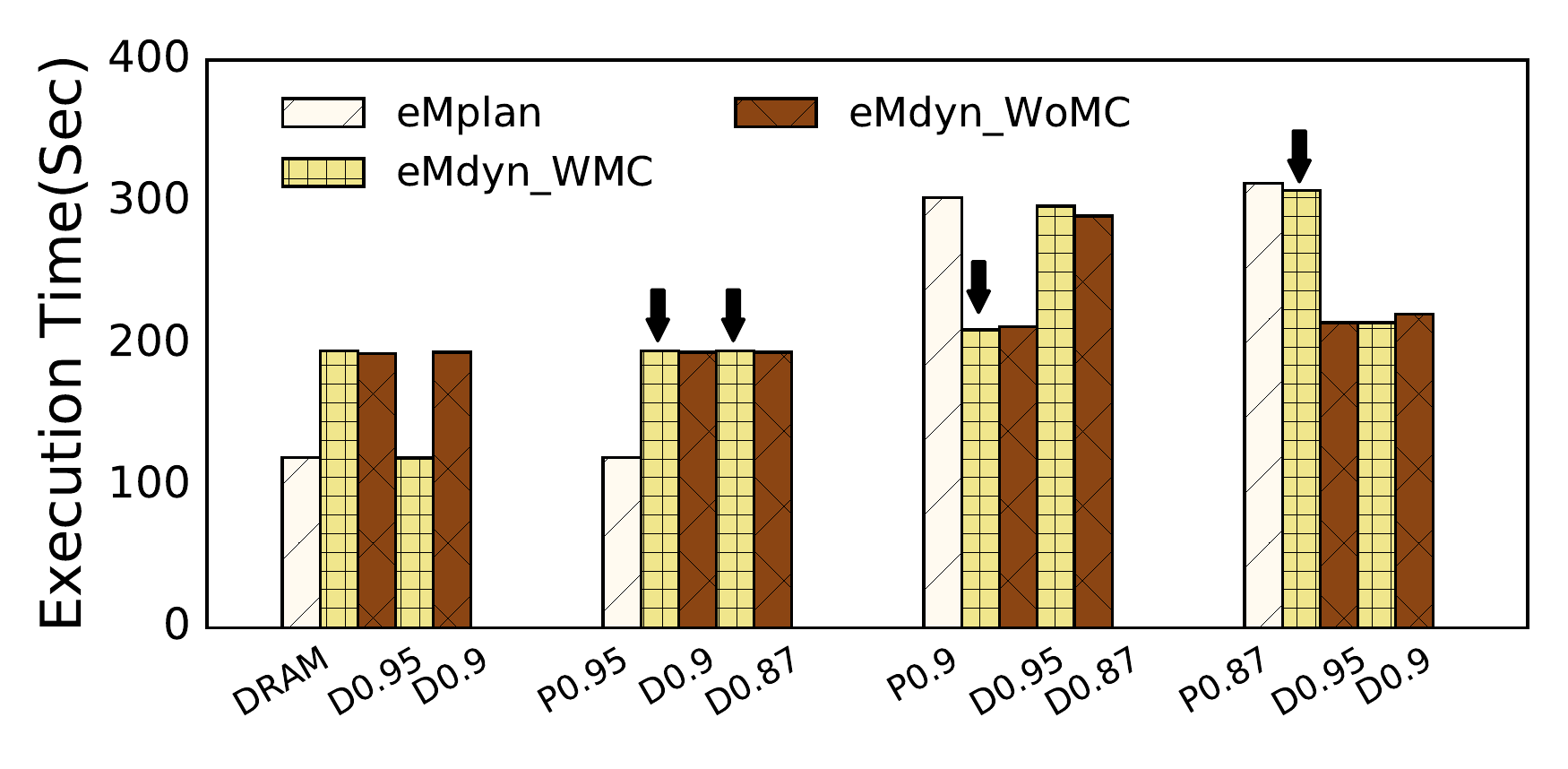}
	} 

	\subfloat[Estimated energy of CG]{
	  \includegraphics[width=0.4\textwidth]{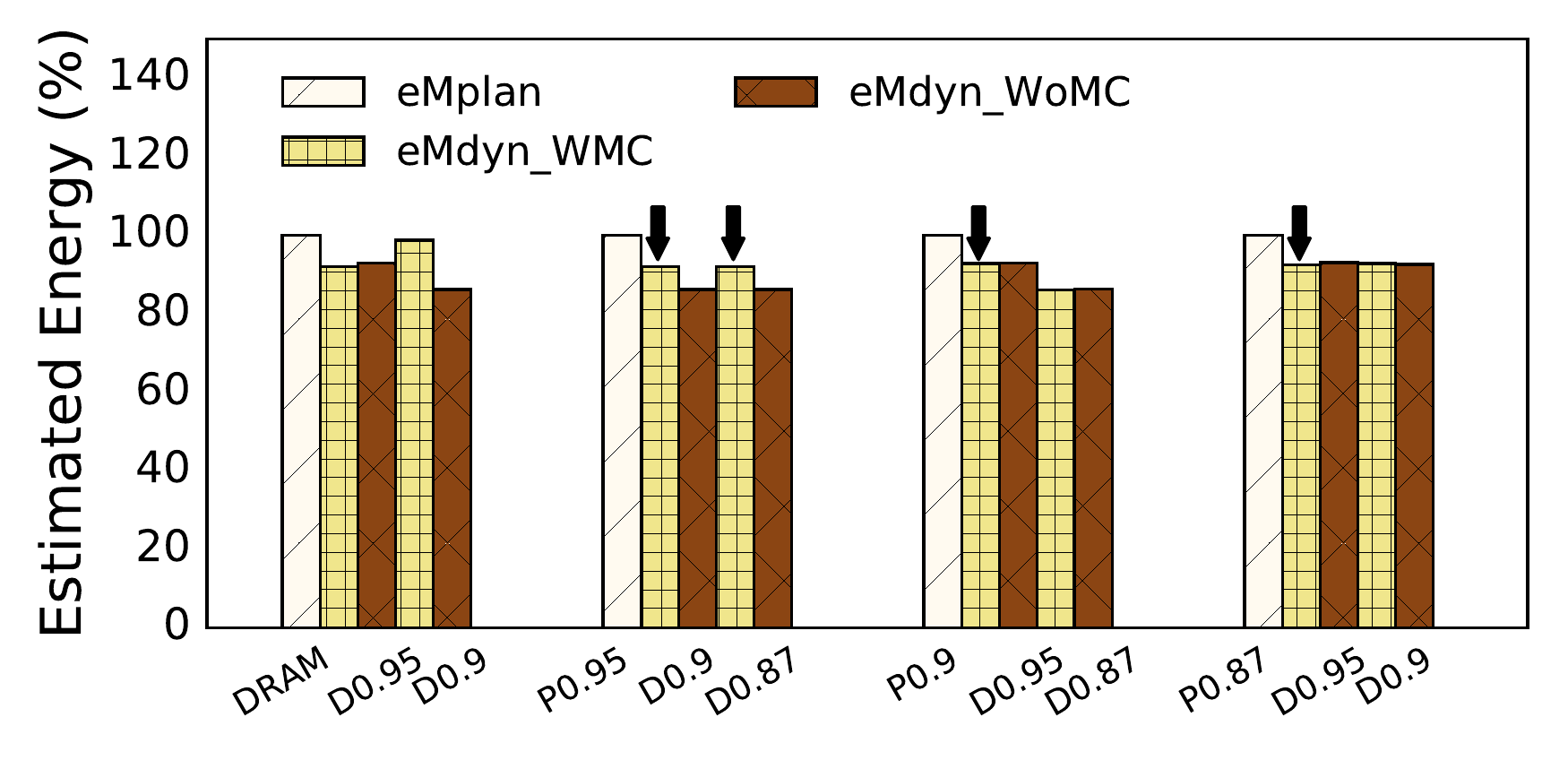}
	}
	\caption{\small The performance and energy consumption of the NPB CG Application. The x-axis is energy constraint where P$x$ shows the energy constraint of \emplan{} as baseline and D$y$ is the changed energy constraint through \emdyn{}.}
	\label{fig:dyn_npb_cg}
\end{figure}

\subsection{\lowercase{e}MDYN Performance and Energy Evaluation} 
In this section, we evaluate the performance and energy efficiency of the second module of \emap{} system, \emdyn{}. Experiments elaborated in this section are performed on Testbed II. We evaluate NPB Benchmark CG and FT applications for \emdyn{} due to their simple code-base and design. We have modified both of the applications to call the member function to register the application pointer addresses as explained in section~\ref{sec:emdyn}.
Due to limited space, we show the evaluation results of only one configuration of HMMS, i.e., HMMS(2,16) where the DRAM capacity is 2 GB and STT-RAM capacity is 16 GB. In the following experiments, we only consider the migration case (1) where an application user deliberately requests for energy efficiency. 
\begin{figure}[t]
	\centering
	\subfloat[Execution time of FT]{
	  \includegraphics[width=0.4\textwidth]{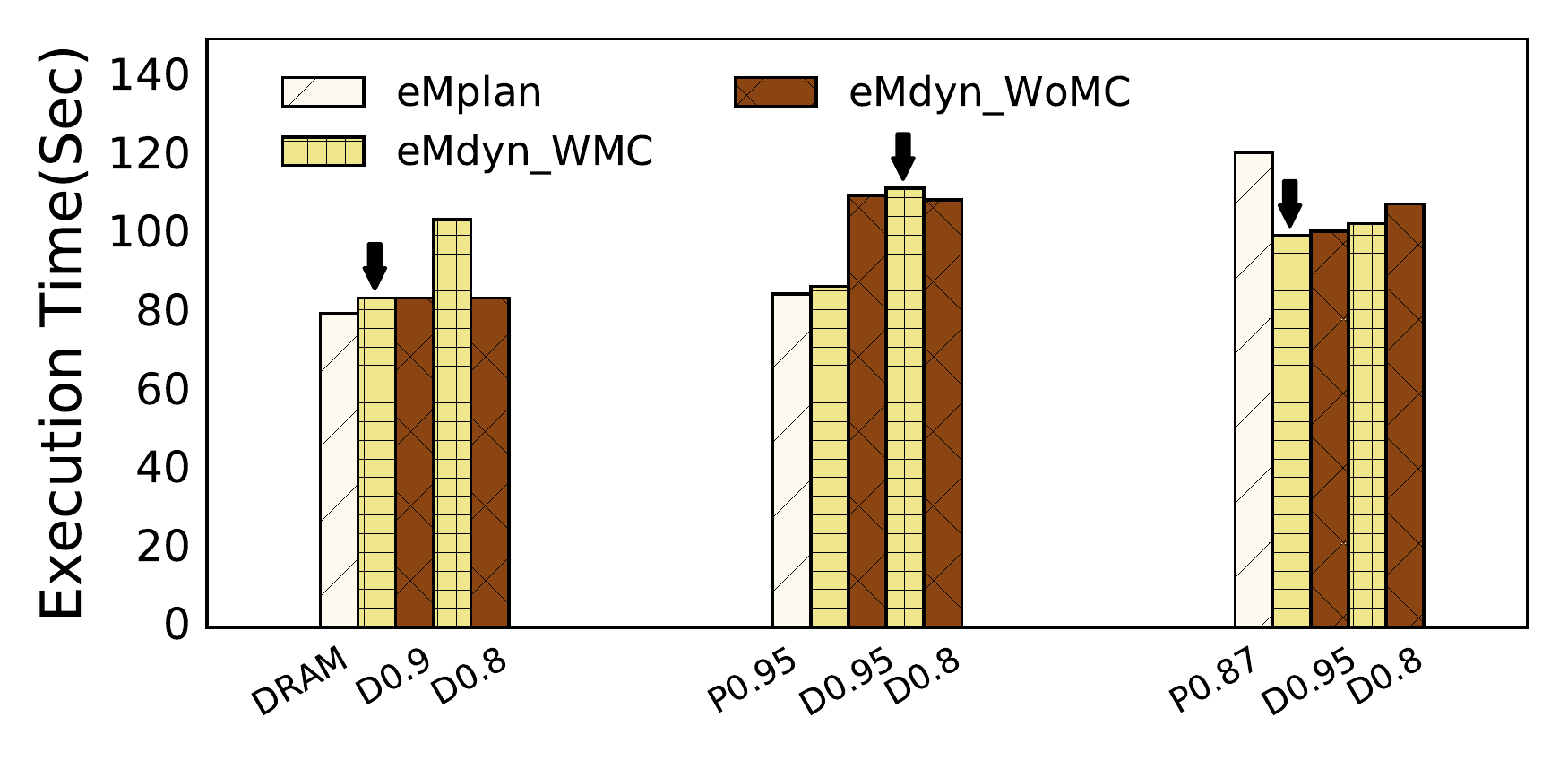}
	} 

	\subfloat[Estimated energy of FT]{
	  \includegraphics[width=0.4\textwidth]{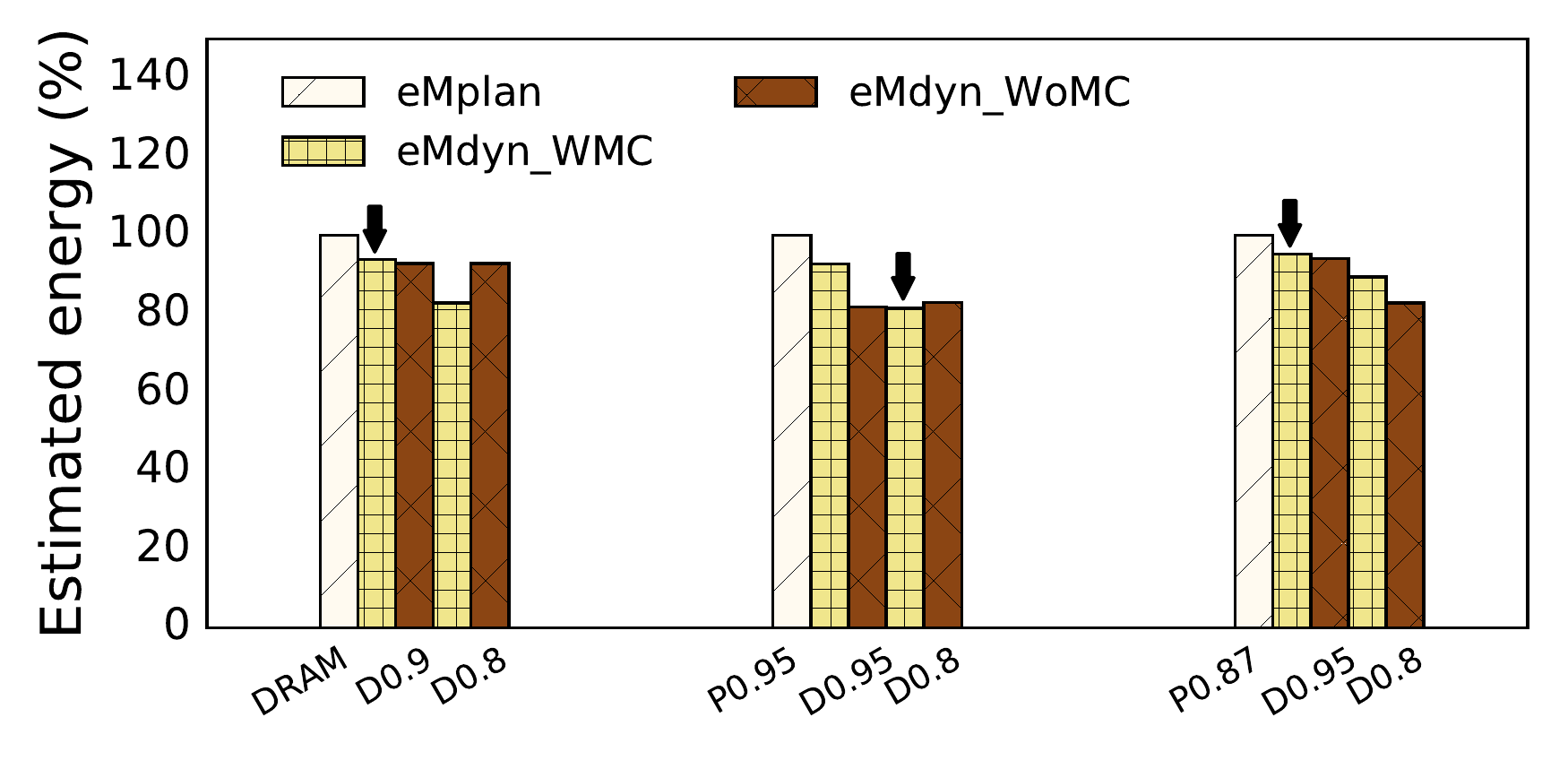}
	}
	\caption{\small The performance and energy consumption of the NPB FT Application. The x-axis is energy constraint where P$x$ shows the energy constraint of \emplan{} as baseline and D$y$ is the changed energy constraint through \emdyn{}.} 
	\label{fig:dyn_npb_ft}
\end{figure}

Figure~\ref{fig:dyn_npb_cg} shows the performance and estimated energy of the CG application under various energy limiting constraints. During the application execution, the request to change the energy limiting constraint occurs and \emdyn{} module is triggered. It re-evaluates the placement of memory objects and shuffles them accordingly. To compute the placement, \emdyn{} interrupts the execution of the application and performs its task and resumes the execution of the application from the same point where it interrupted. In Figure~\ref{fig:dyn_npb_cg}, $eMdyn\_WoMC$ shows the \emdyn{} without considering the migration cost while $eMdyn\_WMC$  is with migration cost. Figure~\ref{fig:dyn_npb_cg}(a) shows that the performance deteriorates as the energy constraint becomes more strict while the performance is improved with week energy constraint. Figure~\ref{fig:dyn_npb_cg}(b) shows that the \emdyn{} reduces the energy consumption as the energy constraint becomes more restricted while the energy consumption is increased if the requested energy limiting constraint is to get more performance. The execution time and the energy consumption of $eMdyn\_WoMC$ is almost similar to the $eMdyn\_WMC$ with energy consumption but at the points shown through arrows in the Figure~\ref{fig:dyn_npb_cg} eMDyn\_WoMC did not meet the performance and energy criteria. This inconsistency of eMDyn\_WoMC is due to not considering the migration cost in terms of energy and performance. 

Figure~\ref{fig:dyn_npb_ft}(a) and (b) show the performance and energy efficiency of \emdyn{} for the FT application. \emdyn{} shows a similar pattern as of CG application. The performance is degraded as the requested energy constraint is more restricted while the energy consumption is reduced. eMdyn\_WoMC also showed a consistent pattern in FT application as CG while the \emdyn{} satisfied the energy and performance in all the cases.

Figure~\ref{fig:perf_break} shows the overall execution time of CG application with \emplan{} and \emdyn{} with various configurations. We modified the CG application and triggered the \emdyn{} on the basis of number of iterations as CG application consists of main loop for computation. We changed the energy limiting constraint during the application execution and the number inside each breakdown of the bar in Figure~\ref{fig:perf_break} shows the changed energy constraint. For the first two bars, we triggered the \emdyn{} after the half number of iterations and show the overhead of \emdyn{}. The other two bars show when the \emdyn{} is triggered after every 20th iteration. From Figure~\ref{fig:perf_break}, it is shown that the overall overhead of \emdyn{} is negligible and it can be called for several times during the execution of the application. But it should be noted that this overhead can be increased according to the number and size of the objects being migrated. 

\section{Related Work}
\label{related}
Various works have been done to optimize the performance and energy efficiency of HMMS through the placement of memory objects. Dullor et al.~\cite{xmem} classify an object into streaming, random, and pointer-chasing patterns based on the dependency and sequentiality of memory access and determine the placement to optimize performance using a greedy algorithm. Wu et al.~\cite{unimem} classified memory objects into bandwidth- and latency-sensitive based on the number of memory accesses and the time taken for the object to optimize performance in MPI applications. However, these works only focus on optimizing performance in the assumption that NVM consumes less power and energy than DRAM. They do not consider that memory energy consumption is affected by the characteristics of NVM devices and object access patterns of application. Further, they also did not consider the energy consumption requirement of various settings.
\begin{figure}[t]
	\centering
	\includegraphics[width=0.38\textwidth]{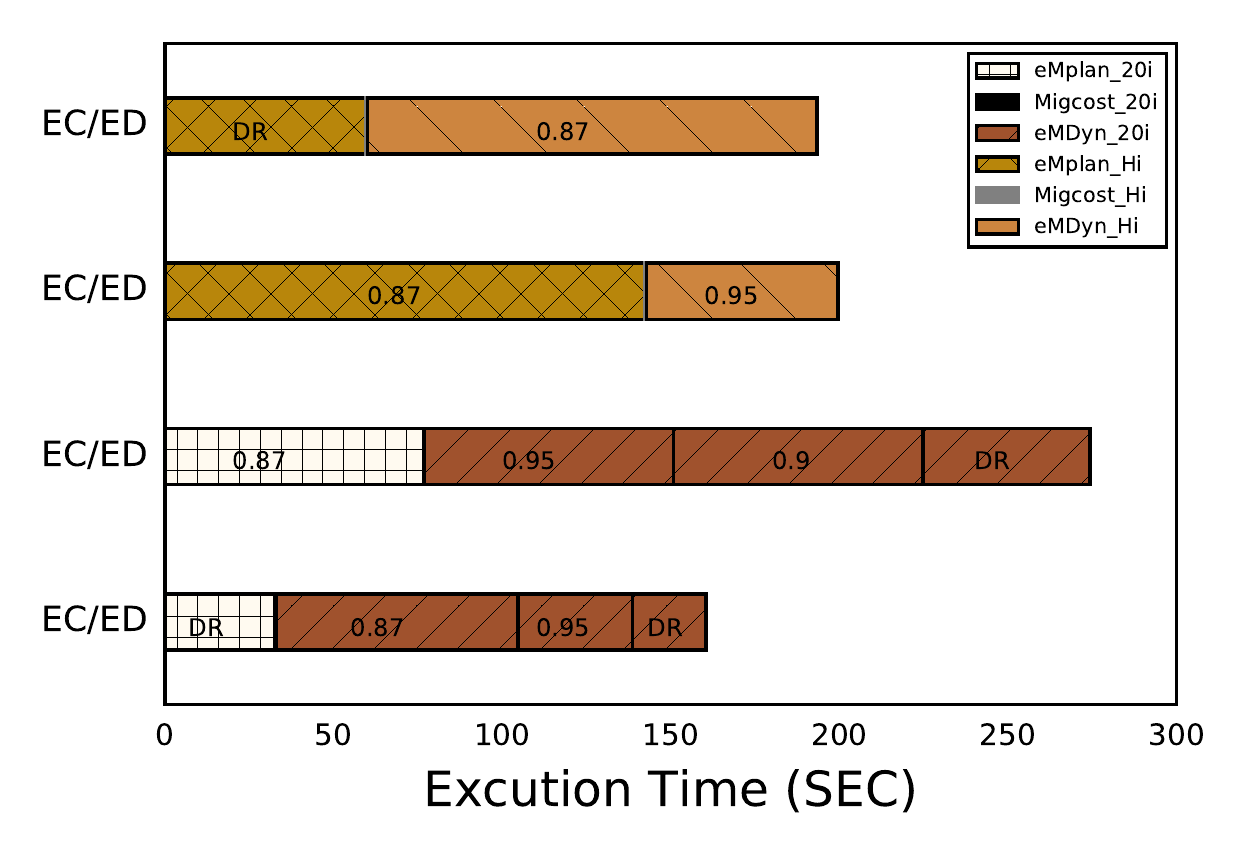}
	\caption{Analysis of time break down of the CG application}
	\label{fig:perf_break}
\end{figure}

In addition, the HMMS which is comprised of high-bandwidth, low-latency, and low-power memory modules is also being studied. MOCA~\cite{moca} and Phadke et al.~\cite{mlpaware} have proposed their solutions for it, which place the object in the most suitable memory device to improve performance and energy efficiency. Phadke et al.~\cite{mlpaware} classified the applications in bandwidth, latency, and power-sensitive and allocates the objects of the application to a best-fit memory module. It only optimizes the performance of the HMMS and does not consider energy efficiency. MOCA~\cite{moca} considers the performance and energy consumption of the ternary HMMS at a finer granularity. They profile the application to obtain the access behavior in terms of LLC MPKI and provide one-time placement of memory objects. MOCA methodology can be applied to binary HMMS consisting of DRAM and NVM. However, MOCA has limitations that it does not estimate the energy consumption by considering the characteristics of the NVM device and the detailed access pattern of the memory object. It also did not take into account the energy requirements during the runtime of the application. 

Existing studies do not consider the amount of energy an object consumes due to its various access patterns and the different characteristics of NVM devices in HMMS. We can optimize the performance and energy efficiency of HMMS through detailed profiling of memory objects access patterns and the NVM device specification. 
\vspace{-0.1in}
\section{Conclusion}
\label{conclusion}

HMMS is a promising solution for an energy-efficient memory system. Albeit, it requires intelligent data placement solutions. Prior solutions either placed application-level or obtained sub-optimal placement of memory objects and only provide static placement schemes. This paper proposed an optimal memory object placement solution by considering both memory access patterns and the nature of memory devices of HMMS. \emap{} calculates the expected energy consumption of objects and allocates the objects to achieve optimal performance, as well as to satisfy the energy limiting constraint. \emap{} provides static (\emplan{}) and dynamic (\emdyn{}) placements of memory objects. \emplan{} places the memory objects at the start of the application by considering their various access patterns and the energy limiting requirements, while \emdyn{} takes into account the changes in energy limiting constraint during the runtime of the application. Our proposed solution meets the energy requirement of 4.17 times less cost while compared to the state-of-the-art memory allocation and classification framework MOCA. It reduces energy consumption by up to 14\% without compromising the performance.

\section*{Acknowledgements}
This research was supported by the Next-Generation Information Computing Development Program through the National Research Foundation of Korea (NRF) funded by the Ministry of Science, ICT (2017M3C4A7080243).

{\bibliographystyle{acm}
\bibliography{ref}}

\end{document}